\documentclass[
 reprint,
 amsmath,amssymb,
 aps,onecolumn,
]{revtex4-2}

\usepackage{graphicx}
\usepackage{dcolumn}
\usepackage{bm}
\usepackage{comment}
\usepackage{mathrsfs} 
\usepackage{tabularx}
\newcolumntype{Y}{>{\centering\arraybackslash}X}
\usepackage{multirow}
\usepackage[ruled,vlined]{algorithm2e}
\usepackage{subfig}
\usepackage{color}

\begin{document}

\preprint{APS/123-QED}

\title{Optimization of a moving sensor trajectory for observing a point scalar source in turbulent flow}

\author{Constantinos~F.~Panagiotou}
 \altaffiliation[corresponding author: ]{co.panayiotou@cut.ac.cy}
\author{Davide~Cerizza}
\affiliation{
 Institute of Industrial Science, The University of Tokyo, Komaba 4-6-1, Meguro-ku, Tokyo 153-8505, 
       Japan
}

\author{Tamer~A.~Zaki}
 \email{t.zaki@jhu.edu}
\affiliation{
 Department of Mechanical Engineering, Johns Hopkins University, Baltimore, MD 21218, USA
}

\author{Yosuke~Hasegawa}
\email{ysk@iis.u-tokyo.ac.jp}
\affiliation{%
 Institute of Industrial Science, The University of Tokyo, Komaba 4-6-1, Meguro-ku, Tokyo 153-8505, 
       Japan
}

\date{\today}
\begin{abstract}
We propose a strategy for optimizing a sensor trajectory in order to estimate 
the time dependence of a localized scalar source in a turbulent enviornment.
We develop and test the algorithm in turbulent channel flow.
The approach leverages the view of the adjoint scalar field as the sensitivity of measurement to 
a possible source. A cost functional is constructed so that the optimal sensor trajectory maintains
a high sensitivity and low temporal variation in the measured signal, for a given source location. 
This naturally leads to the adjoint-of-adjoint equation based on which the sensor trajectory is 
iteratively optimized. It is shown that the estimation accuracy based on the measurement from 
a sensor moving along the optimal trajectory is drastically improved  from that achieved with 
a stationary sensor. It is also shown that the ratio of the fluctuation and the mean of the 
sensitivity for a given sensor trajectory can be used as a diagnostic tool to evaluate the 
resulting performance. Based on this finding, we propose a new cost functional which only 
includes the ratio without any adjustable parameters, and demonstrate its effectiveness in
predicting the time dependence of scalar release from the source.
\end{abstract}

\maketitle

\newcount\ndots
\def\drawline#1#2{\raise 2.5pt\vbox{\hrule width #1pt height #2pt}}
\def\spacce#1{\hskip #1pt}
\def\solid{\drawline{24}{.5}\nobreak\ }
\def\bdash{\hbox{\drawline{4}{.5}\spacce{2}}}
\def\dashed{\bdash\bdash\bdash\bdash\nobreak\ }
\def\bdot{\hbox{\drawline{1}{.5}\spacce{2}}}
\def\dotted{\hbox{\leaders\bdot\hskip 24pt}\nobreak\ }
\def\chndash{\hbox {\drawline{8.5}{.5}\spacce{2}\drawline{3}{.5}\spacce{2}\drawline{8.5}{.5}}\nobreak\ }
\def\chndot{\hbox {\drawline{9.5}{.5}\spacce{2}\drawline{1}{.5}\spacce{2}\drawline{9.5}{.5}}\nobreak\ }
\def\chndotdot{\hbox {\drawline{8}{.5}\spacce{2}\drawline{1}{.5}\spacce{2}\drawline{1}{.5}\spacce{2}\drawline{8}{.5}}\nobreak\ }
\def\chndotdotdot{\hbox {\drawline{8}{.5}\spacce{2}\drawline{1}{.5}\spacce{2}\drawline{1}{.5}\spacce{2}\drawline{1}{.5}\spacce{2}\drawline{8}{.5}}\nobreak\ }
\def\trian{\raise 1.25pt\hbox{$\scriptscriptstyle\triangle$}\nobreak\ }
\def\circle{$\circ$\nobreak\ }
\def\diam{$\diamond$\nobreak\ }
\def\solidcircle{$\bullet$\nobreak\ }

\def\smalltriangle{$\scriptstyle\triangle\textstyle$\nobreak\ }
\def\smallplus{$\scriptstyle + \textstyle$\nobreak\ }
\def\smalltimes{$\scriptstyle\times\textstyle$\nobreak\ }
\def\smallnabla{$\scriptstyle\nabla\textstyle$\nobreak\ }
\def\square{${\vcenter{\hrule height .4pt
        \hbox{\vrule width .4pt height 3pt \kern 3pt
        \vrule width .4pt}
        \hrule height .4pt}}$\nobreak\ }
\def\plus{\raise 1.25pt \hbox{$\scriptscriptstyle +$}\nobreak\ }
\def\x{\raise 1.25pt \hbox{$\scriptscriptstyle \times$}\nobreak\ }
\def\ldash{\hbox {\drawline{7}{.5}\spacce{2}\drawline{7}{.5}\spacce{2}\drawline{7}{.5}}\nobreak\ }
\def\lchndash{\hbox {\drawline{15}{.5}\spacce{3}\drawline{7}{.5}}\nobreak\ }
\def\tsolid{\drawline{24}{1.2}\nobreak\ }

\def\graytrian{\raise 1.25pt
   \hbox to 3bp{
\hfill}\nobreak\ }
\def\trian{\raise 1.25pt
   \hbox to 3bp{
\hfill}\nobreak\ }
\def\solidtrian{\raise 1.25pt
   \hbox to 3bp{
\hfill}\nobreak\ }

\def\graytriand{\raise 1.25pt
   \hbox to 3bp{
\hfill}\nobreak\ }
\def\triand{\raise 1.25pt
   \hbox to 3bp{
\hfill}\nobreak\ }
\def\solidtriand{\raise 1.25pt
   \hbox to 3bp{
\hfill}\nobreak\ }

\def\square{\raise 1.pt
   \hbox to 3bp{
\hfill}\nobreak\ }
\def\graysquare{\raise 1.pt
   \hbox to 3bp{
\hfill}\nobreak\ }
\def\solidsquare{\raise 1.pt
   \hbox to 3bp{
\hfill}\nobreak\ }

\def\circle{\raise 1.pt
   \hbox to 3bp{
\hfill}\nobreak\ }
\def\solidcircle{\raise 1.pt
   \hbox to 3bp{
\hfill}\nobreak\ }
\def\graycircle{\raise 1.pt
   \hbox to 3bp{
\hfill}\nobreak\ }

\def\dotcirc{$\cdots\ $\circle$\cdots$\ }

\def\dashx {\bdash\bdash\smalltimes\bdash\bdash}

\def\chndashx {\drawline{8.5}{.5}\spacce{2}
\drawline{3}{.5}$\scriptstyle\times\textstyle$\drawline{8.5}{.5}\spacce{2}
\drawline{3}{.5}\nobreak\ }

\def\solidcclose{\drawline{10}{.5}\nobreak\raise
  0.5pt\hbox{$\bullet$}\drawline{10}{.5}\nobreak\ }

\def\solidsclose{\drawline{10}{.5}\nobreak\raise
  0.5pt\hbox{\solidsquare}\drawline{10}{.5}\nobreak\ }

\def\solidtclose{\drawline{10}{.5}\nobreak\raise
  0.5pt\hbox{\solidtrian}\drawline{10}{.5}\nobreak\ }

\def\solidcopen{\drawline{10}{.5}\nobreak\raise
  0.5pt\hbox{\circle}\drawline{10}{.5}\nobreak\ }

\def\solidsopen{\drawline{10}{.5}\nobreak\raise
  0.5pt\hbox{\square}\drawline{10}{.5}\nobreak\ }

\def\solidtopen{\drawline{10}{.5}\nobreak\raise
  0.5pt\hbox{\trian}\drawline{10}{.5}\nobreak\ }

\def\solidx{\drawline{10}{.5}\nobreak\raise
  0.5pt\hbox{\x}\drawline{10}{.5}\nobreak\ }

  
\font\msakkk=msam10
\def\diamsol{{\msakkk \char7}}
\def\diamop{{\msakkk \char6}}
\def\starsol{{\msakkk \char70}}
\def\triansolu{{\msakkk \char78}}
\def\triansold{{\msakkk \char72}}
\def\triansolr{{\msakkk \char73}}
\def\triansoll{{\msakkk \char74}}
\def\trianopu{{\msakkk \char77}}
\def\trianopd{{\msakkk \char79}}
\def\trianopr{{\msakkk \char66}}
\def\trianopl{{\msakkk \char67}}
\def\squarsol{{\msakkk \char4}}
\def\squarop{{\msakkk \char3}}

\def\mydash{\hbox{\drawline{2}{.5}\spacce{2}}}
\def\shdashed{\mydash\mydash\mydash\mydash\mydash\mydash\nobreak\ }

\def\bdot{\hbox{\drawline{.5}{.5}\spacce{1}}}
\def\dotted{\hbox{\leaders\bdot\hskip 24pt}\nobreak\ }

\newcommand{\AC}[1]{\vspace{.5cm} [A COMPLETER: #1]}
\newcommand{\CTRClass}{{\it{\bf ctr\_summer.cls }}}
\newcommand{\noi}{\par}


\section{INTRODUCTION}
It is well known that animals and insects heavily rely on their advanced olfactory sense
for a variety of activities, such as food scavenging, mating  and avoiding predators.
As a result, humans  use the superior sensitivity of animals 
in various applications. 
For example, dogs have been extensively used for drug identification and search and rescue operations.
Concurrently, the rapid  advancements in robotics and sensory systems suggest that mobile 
robots or vehicles could successfully replace animals. Such technologies should be 
particularly important for operations under hazardous environments which are inaccessible 
to humans or animals. In order to achieve these goals, the development of efficient algorithms 
for characterizing a scalar source based on limited sensor signals in a turbulent environment 
is important \citep{Kowadlo2008,Ishida12}. 

Existing algorithms for scalar source identification can be classified into two categories:
reactive and model-based algorithms. 
The former prescribes sensor movement based on a sensor signal.
In such a strategy, the flow environment is treated as a black box, and 
the algorithms for sensor movement are often inspired by 
the motions of biological organisms \citep{Kennedy1974,Muller1994,harvey2008}.
In contrast, the latter explicitly takes into account the mathematical models of 
fluid flow and associated scalar transport. Although the computational cost for the latter 
tends to become higher, a model-based method is expected to show superior performance 
than a reactive one, since complex dynamics of scalar transport are incorporated 
in their algorithms. However, their superiority has not been fully verified yet \citep{Voges2014}.

The model-based approaches generally result in an optimization problem, 
in which minimization of a cost functional is required under some constraints.
Optimization techniques can generally be categorized into  probabilistic \citep{Patan2005,Pudykiewicz1998,Sohn2002,Ucinski2000,MonsJCP2019} 
and deterministic methods \citep{Bewley2001,Ucinski2013,WangJCP2019}.
Probabilistic approaches explore a parameter space in a probabilistic way until they reach the global optimum. 
For example, the infotaxis strategy developed by \citet{Vergassola2007} is
based on the Bayesian inference and seeks to minimize the Shannon entropy in order to locate 
the scalar source. \citet{Keats2007} proposed a computationally efficient algorithm 
by combining the adjoint advection-diffusion equation with Monte Carlo Markov Chain sampling 
to evaluate the posterior probability. Although probabilistic approaches are effective in relatively 
simple flow configurations, it is not feasible to extend them to unsteady and three-dimensional 
flow environments, where the degrees of freedom of the design parameters drastically increase.
For example, the algorithm proposed by \citet{Keats2007} requires to solve the adjoint equation 
for each sensor, so that its computational cost increases with a number of sensors.

In the deterministic methods, the set of design parameters are simultaneously
optimized based on the associated gradients of a prescribed cost functional. 
Their unique feature is that the computational cost does not depend on 
the number of design parameters, although a resultant solution may reach a local optimum.
\citet{Cerizza2016}  developed a deterministic method for estimating
the time history of the scalar source intensity at a known source location 
in a fully developed turbulent channel flow reproduced by direct numerical simulation (DNS). 
A finite number of stationary sensors were placed in the downstream of the scalar source, and 
the temporal change of the scalar source intensity was estimated from the downstream sensor signals 
by applying the adjoint analysis. Using a similar approach, \citet{Wang2019} 
reconstructed the spatial location of the source from remote measurements.  
In both studies, a scalar source distribution is estimated by solving the adjoint scalar field, 
which can be interpreted as the sensitivity of a sensor \citep{Keats2007}. 

The adjoint advection-diffusion equation has a source term at the sensor location, and 
it is solved backward in time with a negative advection velocity. Consequently, 
the resultant adjoint scalar field propagates upstream towards possible sources in the past. 
It can be shown that iterative update of an estimated scalar source based on the adjoint scalar field 
guarantees a monotonic decrease in the squared error between the true an estimated scalar concentration 
at a sensor location \citep{Cerizza2016}.  
\citet{Cerizza2016} systematically changed the streamwise distance between a scalar source and a sensor, 
and reported the impact on the estimation performance at different pulsation frequencies of the source.
They also showed that the estimation performance is drastically 
improved with increasing the number of sensors. This suggests that the adjoint scalar fields 
generated at different sensor locations have complementary effects, so that the overall
sensor sensitivity at the source location is enhanced.

Although \citet{Cerizza2016} assumed sensors are stationary, it is expected that higher estimation performance 
could also be achieved by moving a single sensor along a proper trajectory without increasing the number of sensors. 
Indeed, there exist several studies for optimizing sensor arrangement and trajectory.  
In the field of meteorology, the impacts of sensor measurements on the forecast error have been considered
in the framework of static 3D-variational \citep{Baker2000,Langland2004} and 
4D-variational methods \citep{Daescu2008}. The obtained sensitivity information of measurements 
can be used to decide a next sensor location or optimize sensor arrangement \citep{Daescu2008, Misaka2014}.
However, these approaches assume that the grand truth is known, and also require to solve a second-order adjoint
problem, so that a computational cost tends to be large.
In order to overcome the above issues, \citet{Kang2012} extended the concept of unobservability 
proposed by \citet{Krener2009}, and considered the sensitivity of measurements on control variables, e.g., an initial condition 
in the case of 4D variational method. In this framework, a certain sensor arrangement is considered to be better,
if possible changes of control variables could have larger impacts on measurements. This allows to quantitatively 
evaluate a given sensor arrangement without the information on the grand truth. Later, \citet{Mons2017} 
applied optimal control theory to optimization of sensor arrangement for estimating a two-dimensional flow 
past a rotationally oscillating cylinder. A key feature of the latter study is that the adjoint-of-the-adjoint equation
is derived for maximizing the observability, and this opens up a possibility of dealing with control variables
with large degrees of freedom. In the field of informatics, a similar idea of maximizing the sensor sensitivity
on control variables has been proposed by considering Fisher information matrix, and applied to optimization
of the trajectories of mobile sensors \citep{Ucinski2006, Tricaud2010}. However, its application is limited 
to relatively simple problems such as a steady two-dimensional diffusion equation, and also control variables
with limited degrees of freedom.

More recently, data-driven approaches have also been proposed for optimizing sensor arrangement. 
\citet{Verma2020} considered optimization problem of the arrangement of shear-stress and pressure sensors 
around an artificial swimmer to identify the location and the oscillation mode f a nearby object.
They define a utility function which quantify the independence of sensor signals resulting from the object and its
movement, and the sensor locations are optimized so as to maximize the utility function. 
\citet{Deng2021} developed a strategy for
finding minimum sensor locations for identifying the Reynolds-averaged Navier-Stokes (RANS) model constants.
\citet{MonsJCP2019} applied a Kriging-enhanced ensemble variational technique for scalar source estimation,
and obtained the optimal arrangement of multiple stationary sensors by maximizing the condition number
of a matrix describing the source-sensor relationship. Although these approaches have potential for wide applicability,
they commonly require a number of simulations for possible sources (or control variables) before the fact, 
and therefore could be effective only in the cases where the search domain for control variables is confined.

In the present study, we extend the approach based on the sensor sensitivity considered in 
\citet{Kang2012,Mons2017} to trajectory optimization of a moving sensor for observing 
a point scalar source in a fully developed turbulent channel flow.
For this purpose, we revisit the physical meaning of the adjoint scalar field
as the sensitivity of measurement, and formulate the problem as optimization of the sensitivity of a moving
sensor for a given source location. This naturally leads to an extra-adjoint equation, or the
adjoint-of-the-adjoint, based on which the sensor trajectory can be optimized. 
We show that not only the intensity of the sensor sensitivity, but also its uniformity within the search domain is crucial
for better reconstruction of the scalar source. We propose a new cost functional for optimizing the trajectory of 
a moving sensor, and demonstrate that the obtained sensor trajectory results in an estimation 
performance as high as those obtained by seventeen stationary sensors.

The current paper proceeds as follows:
After describing the flow configurations and the numerical scheme in $\S$ 2, we briefly introduce
the estimation results of stationary sensors in $\S$ 3. In $\S$ 4, we develop an optimization 
strategy for a moving sensor trajectory. Then, the estimation performances of a moving sensor 
are presented and compared with those by single and multiple stationary sensors in $\S$ 5. 
Summary and conclusions of the present study are given in $\S$ 6.


\section{COMPUTATIONAL SETUP}\label{numerical_setup}

\begin{figure}
 \centering
  \includegraphics[trim=0cm 0cm 0cm 0cm, clip=true,scale=0.25]{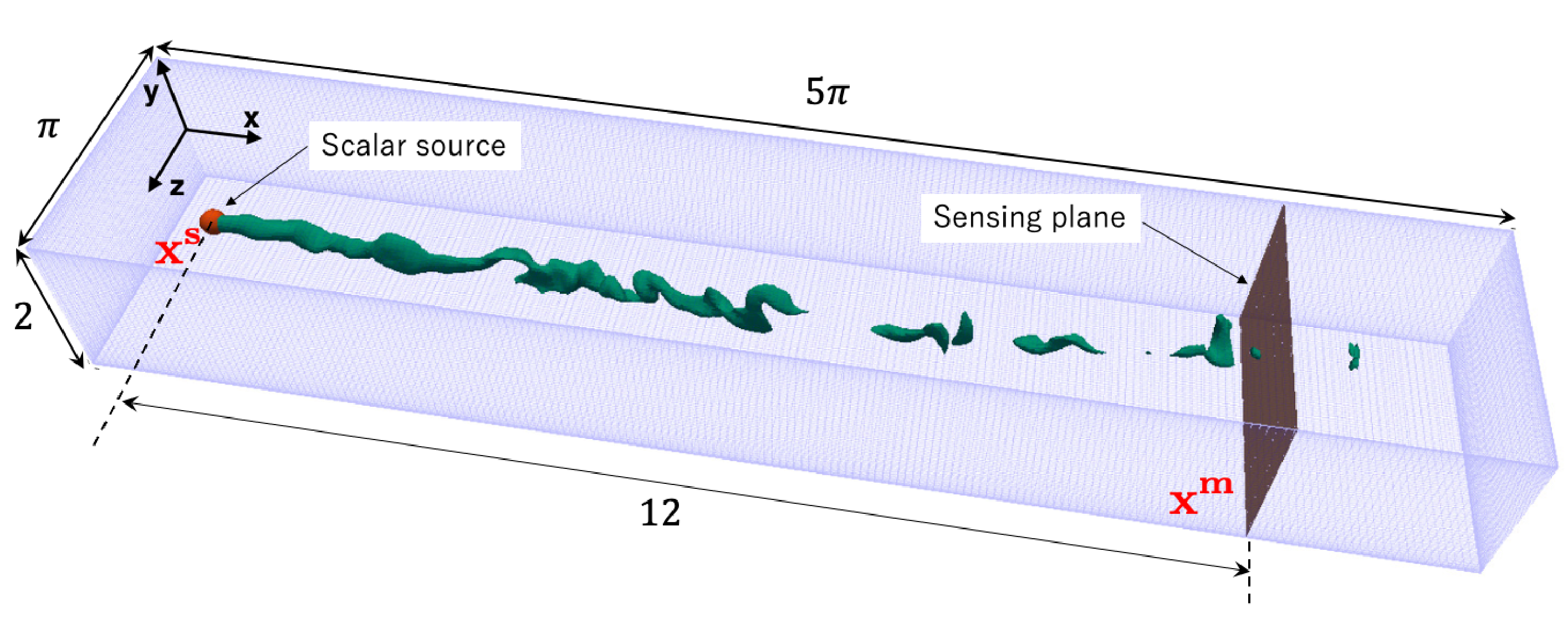}
  \caption{Schematic of the computational domain.}
  \label{fig:computational_domain}
\end{figure}

We consider a statistically stationary turbulent channel flow, in which 
a passive scalar is released from a stationary point source.
Throughout this work, all variables are non-dimensionalized by the friction velocity $u_{\tau}$ and 
the half height of the channel $h$, unless otherwise stated.
Figure~\ref{fig:computational_domain} shows the computational domain and the coordinate system,
where the streamwise,  wall-normal and spanwise coordinates are denoted by $x$, $y$, $z$ respectively. 
The dimensions of the computational domain along the streamwise and spanwise directions are set to be
$5\pi$ and $\pi$, respectively.
The origin of the coordinate system is located at the channel center, so that 
$y = -1$ and $1$ correspond to the locations of the bottom and top walls, 
respectively.
Throughout this study, the position of the scalar source, denoted as $x^{s}_{i}$, 
is located at the centerline of the channel, so that $(x^{s},y^{s},z^{s})=(1.0,0,\pi/2)$.
We assume that the scalar source location is given, while the time history of the scalar source 
intensity is unknown. Hence, our goal is to estimate the scalar source intensity based on 
the sensor measurement downstream. 

The problem design is intended to mimic a hazardous environment where the moving sensor 
can not approach the source, and is hence restricted in its motion.  
The position of a moving sensor is denoted by $x^{m}_{i}(t)$ which is in general an arbitrary
function of time $t$. In the present study, the sensor is assumed to move freely in the $y$-$z$ plane, 
while its $x$-coordinate is fixed to $x^{m}_{1} = 13.0$. 
This way, the estimation performances obtained from stationary and moving sensors will be compared
under the same streamwise separation of $L_{x}=12$ between the source and the sensors.
Hereafter, the $x$-plane where the sensor is located is referred to as a sensing plane. 

We assume an incompressible and Newtonian fluid, so that its governing equations are given 
by the following momentum and continuity equations:
\begin{eqnarray}
\label{eq:NS}
\frac{\partial u_{i} }{ \partial t}+u_{j}\frac{\partial u_{i} }{\partial x_{j} } &=& -\frac{\partial p}{\partial x_{i}}
+\frac{1}{Re_{\tau}}\frac{\partial^{2}u_{i}}{\partial x_{j}^{2}}\,,\\
\label{eq:continuity}
\frac{\partial u_{j} }{ \partial x_{j} } &=& 0,
\end{eqnarray}
where $u_{i}$ is the velocity component in the $i$-th direction and $p$ is the static pressure.
The flow is driven by a constant mean pressure gradient imposed along the streamwise direction $x$,
and the friction Reynolds number is defined as $Re_{\tau}=\frac{ u_{\tau} h}{ \nu }=150$,
where $\nu$ is the kinematic viscosity. This corresponds to the bulk Reynolds number 
 of $Re_{b} \equiv U_b h/ \nu \approx 2300$, where $U_b$ is the bulk mean velocity.
No-slip boundary conditions are imposed at the top and bottom walls,
while periodic conditions are applied in the $x$ and $z$ directions.

The transport equation of a passive scalar $c$ is given by
\begin{equation}\label{forward_c_eq}
\frac{\partial c}{\partial t} + u_{j}\frac{\partial c  }{\partial x_{j} }=
\frac{1}{Pe}\frac{\partial^{2}c }{\partial x_{j}^{2}}+ Q(\bold{x}, t),
\end{equation}
where $Q(\bold{x}, t)$ is a scalar source, which can be considered as 
an arbitrary function of space and time. The Peclet number is defined as $Pe=\frac{ u_{\tau} h }{ \Gamma}$, 
while $\Gamma$ is the coefficient of mass diffusivity of the scalar.
In the present study, the Peclet number is chosen to be equal to $Re_{\tau}$, so that 
the Schmidt  number $Sc = Pe/Re_{\tau} = \nu / \Gamma$ is unity.

Assuming the scalar source is spatially localized and its intensity changes in time 
as $\phi(t)$, the scalar source can be represented by
\begin{equation}
Q(\bold{x}, t)= \phi(t)\delta(\bold{x}-\bold{x}^{s}),
\end{equation}
where $\delta$ is the Dirac's delta function. 
In the current study, we approximate the delta function 
with a steep Gaussian function in order to avoid numerical oscillation,
\begin{equation}
\delta(\bold{x}-\bold{x}^{s}) \simeq  \bigg( \frac{ \beta }{ \pi }\bigg)^{3/2} 
\exp \left( -\beta |\bold{x}-\bold{x}^{s}|^{2} \right)\,,
\end{equation}
 where $\beta$ is chosen so that the scalar source is distributed over several grid points.
The temporal part of the source function $\phi(t)$ is set as\\
\begin{equation}\label{phi_profile}
\phi(t)=\frac{1}{2}\bigg \{ 1 + \cos( 2\pi f t + \pi )\bigg\}\,,
\end{equation}
so that it smoothly changes between $0$ and $1$ at a single frequency $f$.
The initial and boundary conditions for the scalar field are given as
\begin{subequations}
\begin{align}
& c(\bold{x},t=0)=0, \forall \, \bold{x} \in \Omega  \\
& \frac{\partial c }{\partial x_{j}}n_{j}=0 \quad \text{at}\,
y=\pm 1,
\end{align}
\end{subequations}
where $n_j$ represents the outward normal vector on the wall boundary and
$\Omega$ refers to the entire fluid domain. In addition, we remove the scalar in the proximity of the domain boundaries 
along the streamwise and spanwise directions in order to avoid the scalar entering 
from the opposite side by virtue of the periodic conditions

Direct numerical simulation is adopted for solving the velocity and scalar equations. 
For spatial discretization, we use a pseudo-spectral method, where Fourier expansion is adopted 
in the $x$ and $z$ directions, while Chebyshev polynomials are used in the $y$ direction.  
The code has been validated and applied to control and estimation problems in previous studies
\citep{hasegawa2011,suzuki2017}.
The numbers of modes are set to be $(N_1, N_2, N_3) = (128, 65, 64)$ 
along the streamwise, wall-normal and spanwise directions, respectively. 
In order to remove aliasing errors, the 3/2 rule is employed, so that the numbers of 
the physical grid points are 1.5 times larger in all the directions.
The present numerical scheme and condition are essentially identical to those 
in \citet{Cerizza2016}, where extensive verification of the present code is presented.


\section{Scalar source estimation by stationary sensors}

In this section, we first summarize the fundamental mathematical properties of 
the adjoint scalar field, and discuss its physical interpretation
from a viewpoint of scalar source estimation. 
Then, we will show that the adjoint scalar equation can be naturally derived 
by minimizing the squared error between the true and estimated scalar concentration at a sensor location. 
Finally, we present the estimation performances of stationary sensors. These results will
provide us reference data to compare with those by a moving sensor in $\S$ 5.

\subsection{Duality relationship} 

The adjoint scalar field $c^*$ is defined so that it satisfies the following duality relationship:
  \begin{equation}
	  \langle  c^*,   \mathscr{N}\left(c\right) \rangle = \langle  \mathscr{N}^*\left(c^*\right), c \rangle + \mathscr{B}.
	  \label{eq:duality}
 \end{equation}
In equation~(\ref{eq:duality}), the bracket  indicates the spatio-temporal integration, so that 
 \begin{equation}
\langle \,\cdot\, \rangle = \int^{T}_{0} \int_{\Omega}\,\cdot\,dV dt,
\end{equation}
whereas $\mathscr{N}$ and $\mathscr{N}^*$ are an advection-diffusion operator and 
its adjoint, which are respectively defined as
  \begin{eqnarray}
\mathscr{N} = \frac{\partial}{\partial t} + u_j \frac{\partial}{\partial x_j} -  \frac{1}{Pe} \frac{\partial^2}{ \partial x_{j}^{2} },  	  \label{eq:forward_operator}\\
\mathscr{N}^* = -\frac{\partial}{\partial t} - u_j \frac{\partial}{\partial x_j} - \frac{1}{Pe} \frac{\partial^2}{\partial x_{j}^{2} }   	     \label{eq:adjoint_operator}.
 \end{eqnarray}  
The second term on the right-hand-side of Eq.~(\ref{eq:duality}) is a boundary term defined as
\begin{equation}
\mathscr{B} = \langle \frac{\partial \left(c c^*\right)}{\partial t} 
                     + \frac{\partial }{\partial x_j}\left\{c^*cu_j 
                               + \frac{1}{Pe}\left(c\frac{\partial c^*}{\partial x_j} -  c^*\frac{\partial c}{\partial x_j}  \right) \right\}    \rangle.
\end{equation}
Since the integrand has a divergence form, its spatio-temporal integration only depends on the values on the boundaries.
In the present configuration, the boundary term becomes exactly zero, so that we will not consider it hereafter.

Substituting $\mathscr{N}(c) = Q$ and $\mathscr{N}^*(c^*) = \delta(\bold{x}-\bold{x}^{m})\delta(t-t^{m})$ into equation~(\ref{eq:duality}) yields
  \begin{equation}
  \langle c^{*},Q \rangle = \langle \delta(\bold{x}-\bold{x}^{m})\delta(t-t^{m}),c \rangle = c(\bold{x}^{m},t^m) \equiv M(t^m)\,.
  	  \label{eq:duality2}
  \end{equation}
This particular choices of $\mathscr{N}^*(c^*)$ is made, so that the right-hand-side of Eq.~(\ref{eq:duality2}) is equal to
the measurement signal $M\left(t^m\right)$, i.e., the concentration at sensor location $\bold{x}^{m}$ and time $t^m$

Equation~(\ref{eq:duality2}) indicates that the measurement signal can be evaluated in two ways:
We can compute the forward evolution from a source $Q$ and sample at the sensor. Alternatively, we can use
the adjoint field $c^*$ starting from a delta function at the sensor location, and perform dot product with the source $Q$.
A higher value of $c^{*}$ means higher contribution of $Q$ at a particular location and time to the integral, thereby the sensor signal. 
This in turn provides physical interpretation of the adjoint field, that is, the sensitivity of the measurement.

   In contrast to the forward operator~(\ref{eq:forward_operator}) for the scalar field $c$, 
   the adjoint scalar field has to be solved backward in time by using the advection velocity 
   with an opposite sign as can be seen from Eq.~(\ref{eq:adjoint_operator}).
   This is reasonable considering that a sensor signal at a certain location $\bold{x}^{m}$ and time $t^m$
   should be caused by a scalar source present upstream at an earlier time.

\subsection{Algorithm for scalar source estimation} 

In the adjoint-based estimation of scalar source, an arbitrary scalar source distribution
in time and space is first assumed as an initial guess, and then it is iteratively optimized
so as to minimize the error between the estimated scalar concentration and 
the measurement at sensor locations within a certain time horizon $T$.
Hence, we introduce the following cost function $J$:
\begin{equation}\label{cost_func_cerizza}
 J = \int_{0}^{T} \frac{1}{2} \sum_{k} \bigg\{ c(\bold{x}^{m}_{k},t)
     -M_k(t)\bigg\}^{2}dt,
\end{equation}
where  $c(\bold{x}^{m}_{k},t)$ is the estimated scalar concentration at the $k$-th sensor  
location, whereas $M_k(t)$ is the corresponding measurement. In the present study, 
the measurement noise is neglected, so that $M_k(t)$ is directly provided from the DNS results.
The time horizon is set to be $T = 3$, which is normalized by the
friction velocity and the channel half height. This is converted to $T^+ = 450$ in wall units.
The corresponding convection distance $L^+_c = U^+_b \cdot T^+ \approx 6800$ 
is around three times the streamwise dimension of the computational box.

Minimizing $J$ under the constraint of the scalar advection-diffusion equation is 
equivalent to minimizing the following Hamiltonian:
\begin{equation}
\label{eq:Hamiltonian}
H = J- \langle c^{*}\left\{  \mathscr{N} (c)-\phi(t)\delta(\bold{x}-\bold{x}^{s}) \right\} \rangle\,,
\end{equation}
where the adjoint scalar field $c^*$ can be regarded as a Lagrange multiplier.
In the present study, we assume a single point source at a known location $\bold{x}^{s}$.
Hence, our problem is to estimate the temporal change of the source intensity $\phi(t)$.

By applying Fr\'echet differential to $H$ with respect to $\phi(t)$ and 
integration by parts, we finally reach at the following expression (see, \citet{Cerizza2016} 
for the details of the derivation):
\begin{equation}
\label{eq:dhdphi}
 H' \equiv \frac{\mathscr{D}H}{\mathscr{D}\phi} \phi' =  \int^{T}_0 c^*(\bold{x}^{s},t) \phi'(t) dt,
\end{equation}
where a variable with the prime indicates a perturbation caused by an infinitesimal change of $\phi$.
Here, the adjoint scalar field should satisfy the following equation
\begin{eqnarray}\label{adjoint_c_eq}
\mathscr{N}^*(c^*)  &\equiv&
\left(\frac{\partial }{\partial t^{*}} - u_{j}\frac{\partial }{\partial x_{j} } - \frac{1}{Pe}\frac{\partial^{2}}{\partial x_{j}\partial x_{j}}\right)c^* \nonumber\\
&=& \sum_{k}\bigg\{ c(x^{m}_{k},t)-M_{k}(t) \bigg\} \delta(\bold{x}-\bold{x}^{m}_{k})\,,
\end{eqnarray}
under the following initial and boundary conditions:
\begin{subequations}
\begin{align}
&  c^{*}(\bold{x},t^*=0)=0, \forall x \in \Omega, \\
&  \frac{\partial c^{*} }{\partial x_{j}}n_{j}=0 \text{at}\,
y=\pm 1,
\end{align}
\end{subequations}
where the boundary conditions are periodicity in the spanwise $z$-direction, $c^{*}=0$ 
at the inlet and outlet of the channel.
In equation~(\ref{adjoint_c_eq}), $t^{*} = T-t$, so that $t^{*}$ proceeds backward 
in the physical time $t$.

According to equation~(\ref{eq:dhdphi}), the reduction of $H$ is guaranteed when
the source intensity is updated through the following expression:
\begin{equation}
\label{eq:phi_update}
\phi^{n+1}(t)=\phi^{n}(t) -\alpha c^{*}(\bold{x}^{s},t)\,,
\end{equation}
where $\alpha$ is a constant and its optimal value is derived in \citet{Cerizza2016}.
For a detailed discussion regarding the method the reader should refer to the referenced work.
It should be noted that the present algorithm can be applicable to a moving sensor 
by simply assuming that $\bold{x}^{m}$ changes in time. In this case, the location of 
the source term in the adjoint equation~(\ref{adjoint_c_eq}) changes in time accordingly.
The above algorithm will be used for estimating a scalar source intensity 
with a moving sensor in $\S$ 5.

\subsection{Reference source reconstruction from stationary sensors}

\begin{figure}
\centering
\includegraphics[scale=0.25]{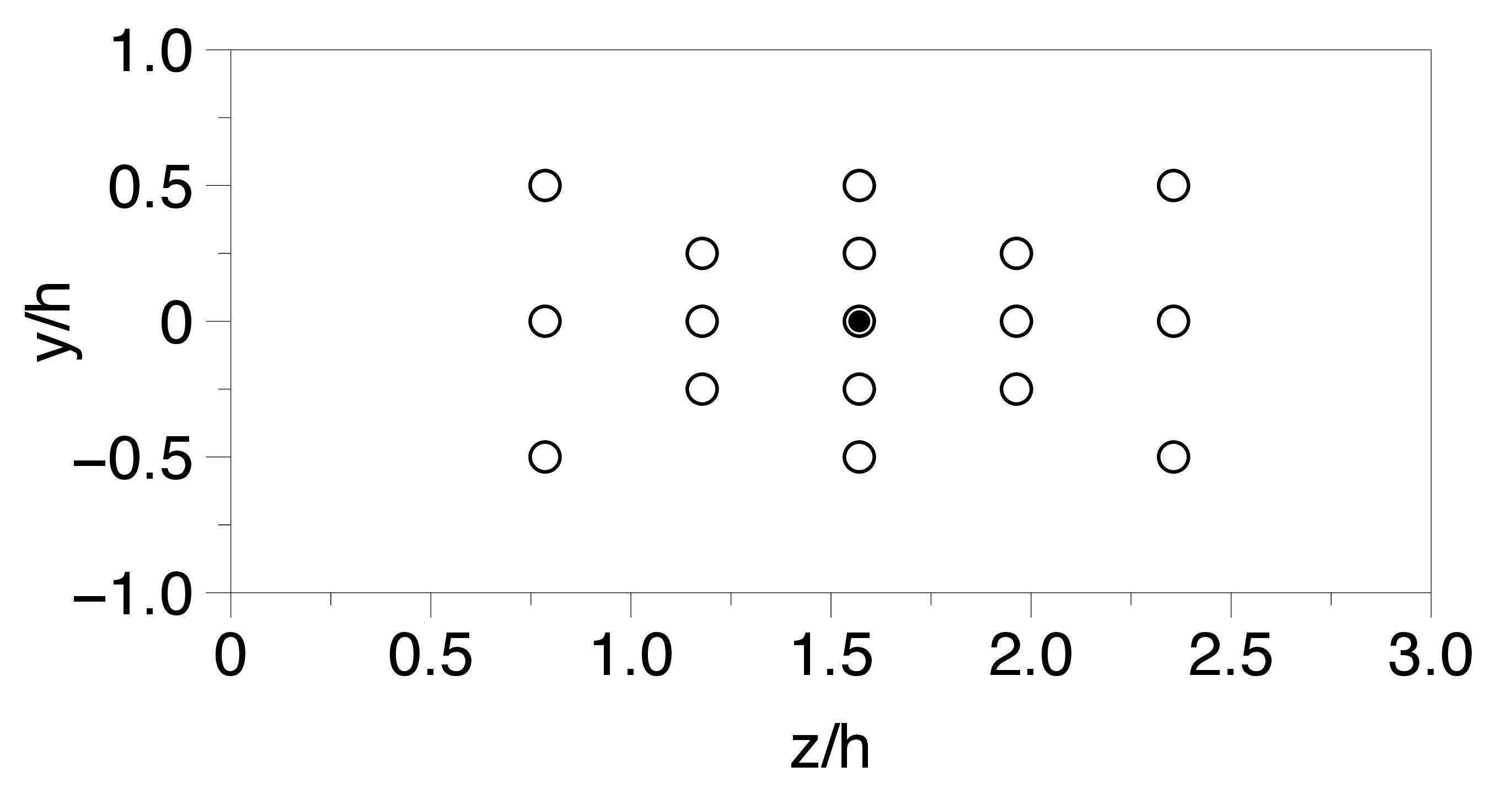} 
\caption{ Sensor arrangement in the $y$-$z$ plane at $x= 13.1$. The open circles 
correspond to the locations of 17 sensors, whereas the filled circle indicates
the location of a single sensor, which is placed at direct downstream of the source.} 
 \label{fig:sensor_arrange}
\end{figure}

In this subsection, we revisit the problem of scalar-source reconstruction 
considered in \citet{Cerizza2016} to get benchmark data, which will be used in $\S$ 5 as 
reference performance.
Following \citet{Cerizza2016}, we placed stationary sensors downstream of the scalar source.
Two different sensor placements are considered as shown in figure~\ref{fig:sensor_arrange}. 
In the first case, a single sensor is placed directly downstream of the scalar source. 
In the second case, seventeen sensors are distributed in the crossflow plane, 
surrounding the central sensor downstream  of the scalar source, 
so that they will cover the mean scalar distribution in the sensing plane.

\begin{figure}
\centering
\includegraphics[scale=0.30]{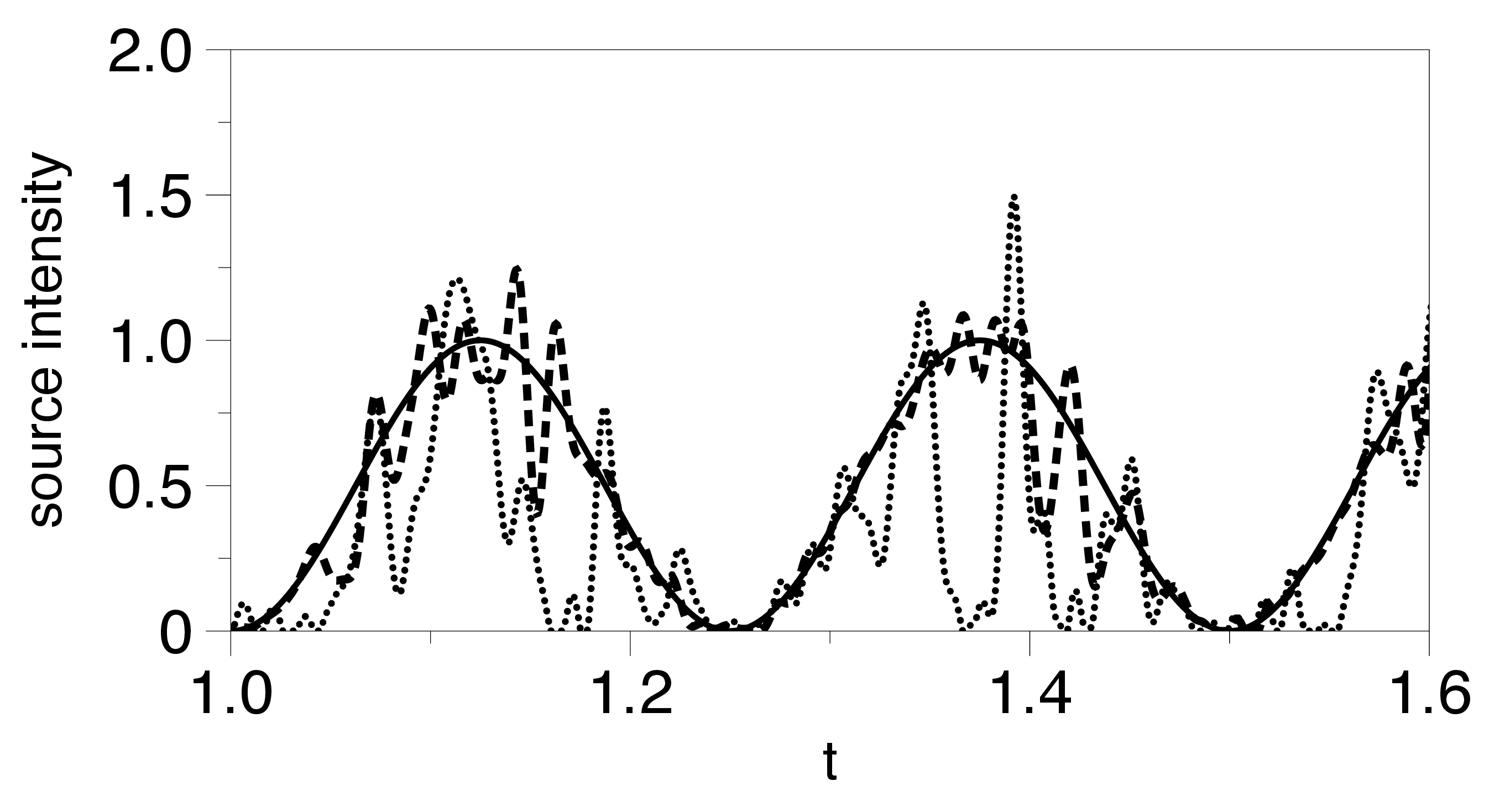} 
\caption{Time trace of source intensity with pulsating frequency $f=4$. 
solid line: true profile, dotted line: estimation with a single sensor, 
dashed line: estimation with seventeen sensors.}
 \label{fig:fixed_multiple_freq2_comp}
\end{figure}

In figure~\ref{fig:fixed_multiple_freq2_comp},  the time traces of the source intensity estimated
from single and seventeen stationary sensors are plotted and compared with the true distribution, 
when the pulsating frequency of the source is set to be $f = 4$.
As anticipated, employing seventeen sensors yields much better estimation 
than that of a single sensor. 
These results suggest that sensor arrangement plays a critical role in 
estimating a scalar source. 

\section{Strategy for optimizing a sensor trajectory}

\subsection{Defining cost-functional}  

The better performance with seventeen stationary sensors shown in the previous section
is due to their collective higher sensitivity to the scalar source location.
This suggests that estimation performance could also be improved with a single sensor 
moving along an optimized trajectory, along which possible scalar releases at any instant can be
captured by the moving sensor.

Suppose that a point scalar source at a location of $\bold{x}^{s}$ instantaneously 
releases scalar at an arbitrary time $t^s$. Such an impulse scalar source may be modeled by 
$Q(\bold{x},t) = \delta(\bold{x}-\bold{x^s})\delta(t - t^s)$.
If we assume continuous measurement in time at a sensor location $\bold{x^m}$,
the time integral of the sensing signal caused by the impulse release can be obtained 
by substituting 
\begin{equation}\label{adjoint_differential_equation}
\mathscr{N}^*(c^*) \equiv \frac{\partial c^*}{\partial t^*} - u_j \frac{\partial c^*}{\partial x_j} - \frac{1}{Pe}\frac{\partial^2 c^*}{\partial x_j \partial x_j} 
= \delta\left(\bold{x} - \bold{x}^{m} \right)\,,
\end{equation}
into the duality relationship~(\ref{eq:duality}). Note that, since we consider the time integral of the measurement signal,
the delta function in time appeared in Eq.~(\ref{eq:duality2}) is removed from $\mathscr{N}^*(c^*)$ here.
Indeed, the above choice of $\mathscr{N}^*(c^*)$ leads to
\begin{equation}
\tilde{M} \equiv \int_{-\infty}^{\infty} M(t) dt = \left< c^*, \delta(\bold{x}-\bold{x^s})\delta(t - t^s) \right> = 
c^*\left( \bold{x^s}, t^s\right),
\end{equation}
where a tilde represents an integral over time.
The above equation indicates that 
the time integral of measurement at the sensor location $\bold{x^m}$ can be given by
the instantaneous adjoint scalar concentration at the source location $\bold{x^s}$ and
the instant of scalar release $t^s$.

Here, we make the following assumptions for better performance of scalar source estimation:
First, when estimating scalar release from a source at a certain time $t^s$, the resultant time integral of 
the sensor signal should be high. Indeed, if the scalar plume does not overlap the sensor location and 
there is no sensing signal, it is impossible to reconstruct the source. Second, the sensing signal resulting from
an instantaneous scalar release within the time horizon should have similar intensity regardless of releasing time. 
If the sensor signal is weak for a particular releasing time, namely, the sensitivity of the sensor is low for the instantaneous scalar release,
it becomes more difficult to accurately reconstruct the scalar source at the instant. For example, the source reconstruction
with a single sensor presented in Fig.~\ref{fig:fixed_multiple_freq2_comp} shows large fluctuation, which may be 
attributed to the fluctuation of the sensitivity for instantaneous scalar release. It should be noted that the first assumption
is similar to those adopted in \citet{Kang2012,Mons2017}, where the sensor arrangement is optimized so as to maximize the magnitude of the sensor
sensitivity. Meanwhile, the second assumption, i.e., the uniformity of the sensitivity within the search domain, is newly introduced 
in the present study. As will be shown later, the second assumption further improves the estimation accuracy.

Based on the above considerations, we aim to maximize $c^*$ at the source location
with less fluctuation in time, and define the following cost functional
to be minimized for finding the optimal sensor trajectory:

\begin{equation}\label{cost_fun_or}
J_{1} = - \alpha_{1}\int_{0}^T c^*(\bold{x}^s, t) dt + \alpha_{2}\int_{0}^T 
\bigg( c^{*}(\bold{x}^s, t)-\overline{c^{*}}(\bold{x}^s)\bigg)^{2}dt 
+\alpha_{3}\int_{0}^T \bold{u}^{m}\cdot\bold{u}^{m}dt, 
\end{equation}
where $\bold{u}^{m}=\frac{d \bold{x}^{m} }{dt}$ denotes the sensor velocity.
The positive weighting constants $\alpha_{1}$, $\alpha_{2}$ and $\alpha_{3}$ are
introduced in order to change the relative contribution from each term to $J_{1}$. 
The first term contributes to increase the averaged sensitivity at the source location, 
while the second term tends to reduce its temporal fluctuation.
The third term represents a penalty on the sensor speed, and works as a regularization term. 
The time horizon is set to $T = 3$, which is identical to that used for scalar source 
estimation from a stationary sensor in $\S$~3.

 \subsection{Parameters for optimizing a sensor trajectory }
 
In order to systematically change the contributions from the three terms in $J_1$, 
we introduce the following two parameters. 

The first one is 
 \begin{equation}\label{R_ratio}
R_{21}= \frac{ \alpha_{2} \int_{0}^T \bigg[  c^{*}(\bold{x}^{s},t)-\overline{c^{*}}(\bold{x}^{s})\bigg]^{2}dt }
{ \alpha_{1} \int_{0}^T c^{*}(\bold{x}^{s},t)dt  }\,, 
\end{equation}
which represents the ratio of the second and first terms in $J_1$.
The weighted constants $\alpha_{1}$ and $\alpha_{2}$ are determined so as to realize 
a prescribed value of $R_{12}$. 
A large value of $R_{21}$ leads to a larger weight for the second term in $J_1$ suppressing 
the temporal fluctuation of $c^*$ at the source location than the first term for maximizing 
the averaged value of $c^*$. The second parameter $R_{31}$ corresponds to the ratio of
the third and first terms of $J_1$, and is defined as
\begin{equation}\label{R_ratio2}
R_{31}= \frac{  \alpha_{3}\int_{0}^T u^{m}_{j} u^{m}_{j}dt  }
{  \alpha_{1}\int_{0}^T c^*(\bold{x}^s, t) dt   }.
\end{equation}
Again, $\alpha_{1}$ and $\alpha_{3}$ will be adjusted so as to achieve a prescribed value of $R_{31}$. 

Table~\ref{table:global_matrix} summarizes all combinations of $R_{21}$ and $R_{31}$ 
considered in the present study. In Case $A$, both $R_{21}$ and $R_{31}$ are null, so
that only the first term of $J_1$ is taken into account. On the other hand, for cases $B_1$-$B_4$, 
the weight for the second term is systematically increased while the third term is kept zero. 
Finally, in case $C$, the effects of the third term are examined.
Note that the time integrals in equations~(\ref{R_ratio}, \ref{R_ratio2}) are in general unknown before
optimizing a sensor trajectory. In the present study, their values are evaluated from the statistics of 
$c^*$ for Case $A$, so that the values of $\alpha_{1}$, $\alpha_{2}$ and $\alpha_{3}$ are 
all fixed during the procedures of optimizing a sensor trajectory.

\subsection{Performance indices}
Once a sensor trajectory is successfully optimized in Cases B and C,
it should reduce the temporal fluctuation of $c^*$ with respect to its temporal average
$\overline{c^*}$ according to the definition of $J_1$ in equation~(\ref{cost_fun_or}). 
This motivates us to define the following ratio of the standard deviation and the time average of $c^*$
at the source location:
\begin{equation}\label{epsilon_rms_mean}
\epsilon = \frac{ \sqrt{ \frac{1}{T} \int_{0}^T\bigg( c^{*}(\bold{x}^s, t)-
\overline{c^{*}}(\bold{x}^s)\bigg)^{2}dt} } 
{ \frac{1}{T} \int_{0}^T c^*(\bold{x}^s, t) dt },
\end{equation}
which will be used to evaluate different sensor trajectories later.

As for a performance index of scalar source estimation,
we define the correlation coefficient $\psi^{\phi}$ between 
the time traces of the true and estimated source intensities
i.e., $\phi^{true}(t)$ and $\phi^{est}(t)$ as
\begin{equation}
\psi^{\phi}=\frac{ \int_{0}^T (\phi^{true}(t) -\overline{\phi^{true}} ) (\phi^{est}(t) -\overline{\phi^{est}} )dt  }
{\sqrt{  \int_{0}^T (\phi^{true}(t) -\overline{\phi^{true}} )^{2}dt} \sqrt{ \int_{0}^T (\phi^{est}(t) -\overline{\phi^{est}} )^{2}dt} }\,.
\end{equation} 
Obviously, $\psi^{\phi}$ closer to unity indicates a better performance.

Another index is the 
L2 norm of the difference between $\phi^{true}(t)$ and $\phi^{est}(t)$,
given by
\begin{equation}
\ell^{2}_{norm}=\sqrt{ \frac{1}{T}\int_{0}^T 
\bigg(\phi^{true}(t)-\phi^{est}(t)\bigg)^{2} dt }\,.
\end{equation}
 A smaller value of $\ell^{2}_{norm}$ means a less estimation error, 
 and thereby better estimation. In the following sections, 
 $\psi^{\phi}$ and $\ell^{2}_{norm}$ will be used to evaluate 
 the estimation performances in all cases.

\begin{table}
\centering
\begin{tabularx}{\textwidth}{@{}lYY|YYY|YYYY  @{}}
\multirow{2}{*}{Case} &  \multirow{2}{*}{$R_{21}$}& \multirow{2}{*}{$R_{31}$} &            \multicolumn{3}{ c }{circular}               &     \multicolumn{3}{ c }{random} \\ 
                      &                           &                           & $\epsilon$         & $\psi^{\phi}$  &  $\ell^{2}_{norm}$  & $\epsilon$      & $\psi^{\phi}$  &  $\ell^{2}_{norm}$  \\ 
    $A$               &    0.0                    &    0.0                    &    1.23            & 0.74           &         0.29        &     1.10        &      0.80      &    0.25             \\ 
    $B_{1}$           &    0.1                    &    0.0                    &    1.18            & 0.76           &         0.27        &     1.06        &      0.82      &    0.24             \\ 
    $B_{2}$           &    1.0                    &    0.0                    &    0.94            & 0.77           &         0.26        &     0.85        &      0.89      &    0.17             \\ 
    $B_{3}$           &    10                     &    0.0                    &    1.30            & 0.55           &         0.39        &     1.17        &      0.58      &    0.36             \\ 
    $B_{4}$           &    $\infty$               &    0.0                    &    1.39            & 0.53           &         0.40        &     1.25        &      0.55      &    0.37             \\ 
    $C$               &    1.0                    &    1.0                    &    1.03            & 0.69           &         0.31        &     0.97        &      0.74      &    0.28             \\ 
\multicolumn{3}{ c| }{initial}                                                 &    1.82            & 0.43           &         0.46        &     1.64        &      0.52      &    0.42               \\ 
\end{tabularx}
\caption{ Summary of the numerical conditions and the resultant estimation performance.
}
\label{table:global_matrix}
\end{table}

\subsection{ Derivation of optimization algorithm  }
\label{optimization_strategy}

Minimizing $J_{1}$ given by equation~(\ref{cost_fun_or}) requires optimization of the adjoint scalar field
which is governed by equation~(\ref{adjoint_differential_equation}). Hence, we define the following
Hamiltonian for optimizing a sensor trajectory:

\begin{equation}\label{hamiltonian_exp}
H = J_{1} + \left< \theta \left\{ \frac{\partial c^*}{\partial t^*} - u_j \frac{\partial c^*}{\partial x_j} 
- \frac{1}{Pe}\frac{\partial^2 c^*}{\partial x_j \partial x_j} 
- \delta\left(\bold{x} - \bold{x}^{m} \right) \right\} \right>,
\end{equation}
where $\theta$ is a Lagrange multiplier, and can also be considered as 
the adjoint of the adjoint field. Our purpose is to optimize the sensor trajectory
$\bold{x}^{m}(t)$ to minimize $H$. For this purpose, we can take a standard approach,
which is essentially the same as equations~(\ref{eq:Hamiltonian})-(\ref{eq:phi_update})
used for estimating scalar source. Namely, applying the Fr\'echet differential to 
equation~(\ref{hamiltonian_exp}) with respect to an infinitesimal change $(\bold{x}^{m})'$
of a sensor trajectory, and integrating by parts (see, Appendix~A for more details), we end up with

\begin{eqnarray}\label{simplified_Hvar}
 H' \equiv \frac{\mathscr{D}H}{\mathscr{D}\bold{x}^{m}} (\bold{x}^{m})'
&=& \left< -\frac{\partial \theta}{\partial x_j} \delta\left({x}_j - {x}^{m}_j\right) (x^{m}_j)' \right>- 
 2 \alpha_{3} \int \frac{ d^{2} x^{m}_{j} }{ dt^{2} } (x^{m}_j)' dt \nonumber\\
&=& \int_{0}^{T}  -\bigg[ \left( \frac{\partial \theta}{\partial x_j} \right)_{ \bold{x}^{m} }
+2 \alpha_{3}\frac{ d^{2}x^{m}_{j} }{ dt^{2} }\bigg] (x^{m}_j)'  dt, 
\end{eqnarray}
where $\theta$ has to satisfy the following equation:
\begin{equation}\label{theta_evol_eq}
\frac{\partial \theta}{\partial t} + u_{j}\frac{\partial \theta}{\partial x_{j}}
=\frac{1}{Pe} \frac{\partial^2 \theta}{\partial x_{j}^{2} }
+ \bigg\{\alpha_{1} - 2 \alpha_{2}\left(c^{*}-\overline{c^{*}}\right)\bigg\} \delta(\bold{x}-\bold{x}^{s}),
\end{equation}
with the following initial and boundary conditions
\begin{eqnarray}\label{theta_bc}
\theta(\bold{x},t=0)=0,\quad \forall \, \bold{x} \in \Omega \\
\frac{\partial \theta }{\partial x_{j}}n_{j}=0,\quad \text{at}\,
\partial \Omega.
\end{eqnarray}

Equation~(\ref{simplified_Hvar}) indicates that $H'$ is always negative 
if the sensor trajectory is updated based on the following expression:
\begin{equation}\label{update_exp}
(x^{m}_j)' (t) \equiv x^{m,n+1}_{j}(t)-x^{m,n}_{j}(t) =\alpha^{n}
\bigg[ \left( \frac{\partial \theta}{\partial x_j} \right)_{ \bold{x}^{m} }
+2 \alpha_{3}\frac{ d^{2} x^{m}_{j} }{ dt^{2} }\bigg]^{n},
\end{equation}
where the superscript $n$ indicates the current iteration step and 
$\alpha$ is a positive coefficient determining the amount of the update
in each iteration step.

\begin{algorithm}[H]
\SetAlgoLined
\label{algorithm}
 \caption{Algorithm for optimizing sensor trajectory}
 \flushleft
\begin{itemize}
\item $n=0$;
\item Solve equations~(\ref{eq:NS}, \ref{eq:continuity}) and store all data regarding the
 velocity field; \\
\item Prescribe an initial guess of the sensor trajectory $\bold{x}^{m,0}$; \\
\end{itemize}
 \While{Convergence criterion is not satisfied}{
  \begin{itemize}
  \item Advance the adjoint equation $\mathscr{N}^*(c^*)=\delta\left(\bold{x}-\bold{x}^{m}\right)$
   and record $c^{*}$ at the source location to evaluate 
   $\overline{c^{*}}=\frac{1}{T}\int^{T}_{0}c^{*}(\bold{x}^{s},t)dt$;
  \item Advance the adjoint-of-adjoint equation~(\ref{theta_evol_eq}) for $\theta$ 
  and store $\frac{\partial \theta}{ \partial x_{j} }$ at sensor location for the entire time period;
  \item Compute step size, $\alpha^{n} = 0.05/\max \bigg| \bigg[\bigg( 
  \frac{\partial \theta}{ \partial x_{j} } \bigg)_{\bold{x}^{m}} + 
  2\alpha_{3}\frac{d^{2} x^{m}_{j} }{dt^{2}}\bigg]^{n}  \bigg|$;
  \item Update sensor's trajectory in accordance with equation~(\ref{update_exp});
  \item $n=n+1$  
  \end{itemize}
 }
\end{algorithm}

The overall procedure for optimizing a sensor trajectory is summarized in algorithm~\ref{algorithm}.
First, we solve equations~(\ref{eq:NS}, \ref{eq:continuity}) and store 
the spatio-temporal evolution of the entire velocity field
within the computational domain $\Omega$ and the time horizon $0 \le t \le T$.
This has to be done only once before starting optimization, and
the same data set of the velocity field can be used throughout 
the following optimization procedures.
Secondly, assuming an arbitrary initial sensor trajectory $\bold{x}^{m, 0}$, 
the adjoint equation~(\ref{adjoint_differential_equation}) is solved backward in time
and $c^*$ at the source location is recorded. Then, this information is used to solve the 
extra-adjoint equation~(\ref{theta_evol_eq}) for $\theta$. Note that, in contrast to 
equation~(\ref{adjoint_differential_equation}), equation~(\ref{theta_evol_eq}) is solved forward in the original time $t$.
Once $\theta$ at the sensor location is obtained, the sensor trajectory is updated in accordance with 
equation~(\ref{update_exp}). The above procedures are repeated until the sensor trajectory converges.

\begin{figure}
\centering
\includegraphics[trim={1cm 6cm 0 6cm},clip,scale=0.7]{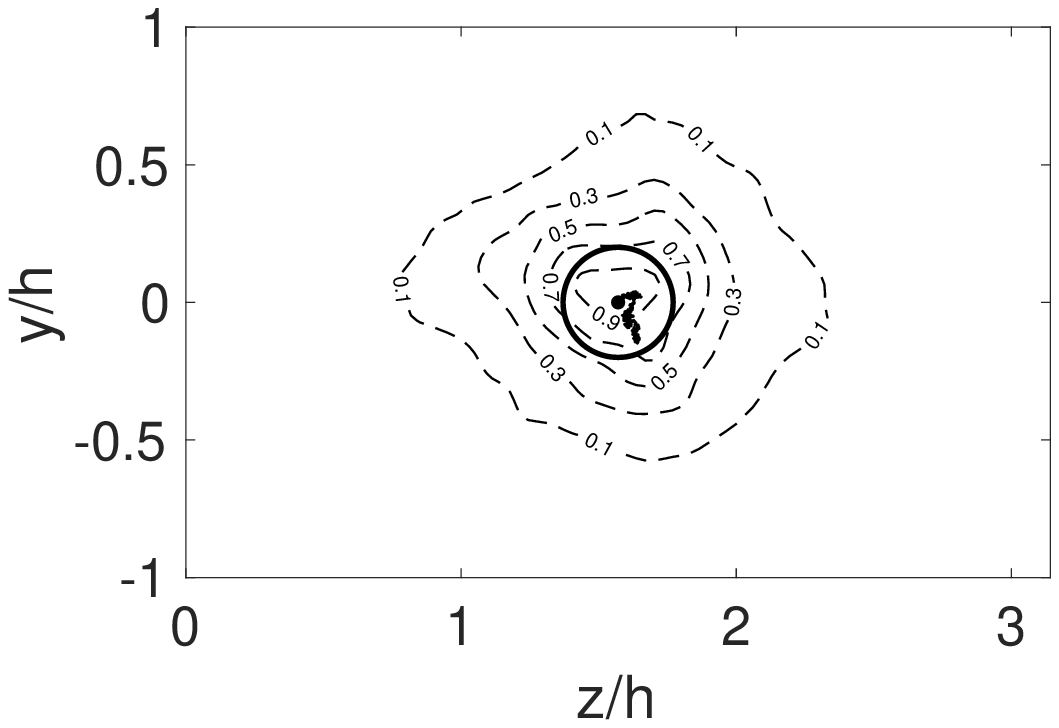} 
\caption{Isolines of the averaged  scalar field  $c$ at the $y$-$z$ plane of motion,
 where  the scalar field is time-averaged within the time interval $t \in (0,T)$.
 The values of the scalar field are normalized to the  maximum  averaged value.
 Solid lines indicate  the initial sensor trajectories.  } 
 \label{fig:contour_theta}
\end{figure}

We considered two different initial trajectories which are both lying sufficiently close
to the core of the plume at the measurement plane, so that a sensor receives significant signals. 
The first one is 
a circular motion, the center of which is located at the channel center and its radius is $0.2$, and
the second one is a trajectory generated by a random walk starting from the channel center.
These two initial trajectories are depicted in figure~\ref{fig:contour_theta},
in which the iso-lines of the mean scalar concentration from the point source at $\bold{x}^{s}$ 
with a steady release are also shown. It can be confirmed that both initial trajectories
are within these contours.

\section{Results}
We first show the optimal sensor trajectories obtained for Cases $A$-$C$ in $\S$ 5.1.
In $\S$ 5.2, the estimation performances based on the signals obtained from the optimal sensor trajectories
are presented and compared with those of stationary sensors. Based on the obtained
results, we propose a simpler and more effective strategy for optimizing a moving sensor trajectory, 
and validate its performance in $\S$ 5.3.

\subsection{Optimal sensor trajectory}  
\subsubsection{Cases $A$ and $B$}
\begin{figure}
\flushleft
\subfloat[]
{\label{}
\includegraphics[width=0.5\textwidth]{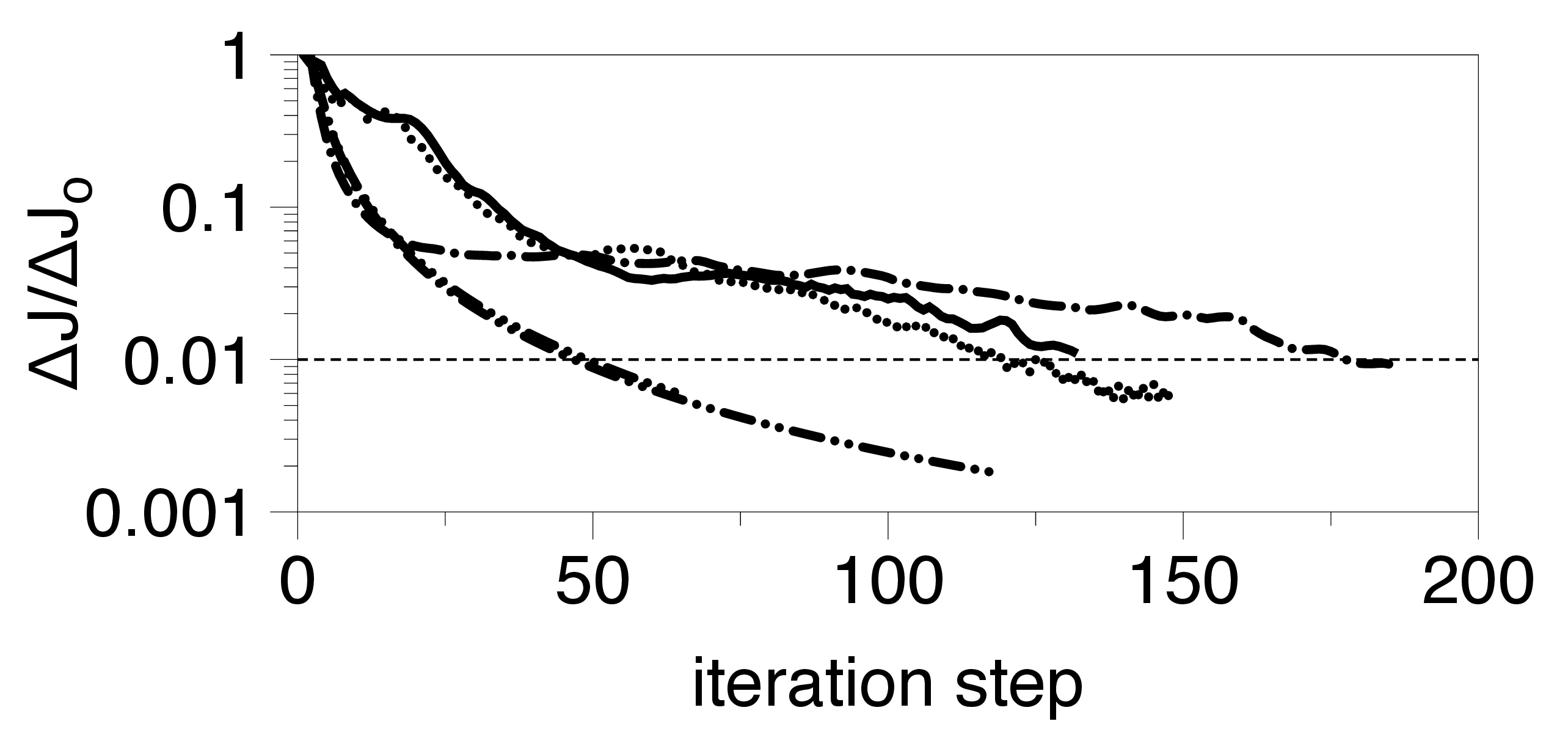}} 
\subfloat[ ]
{\label{}
\includegraphics[width=0.5\textwidth]{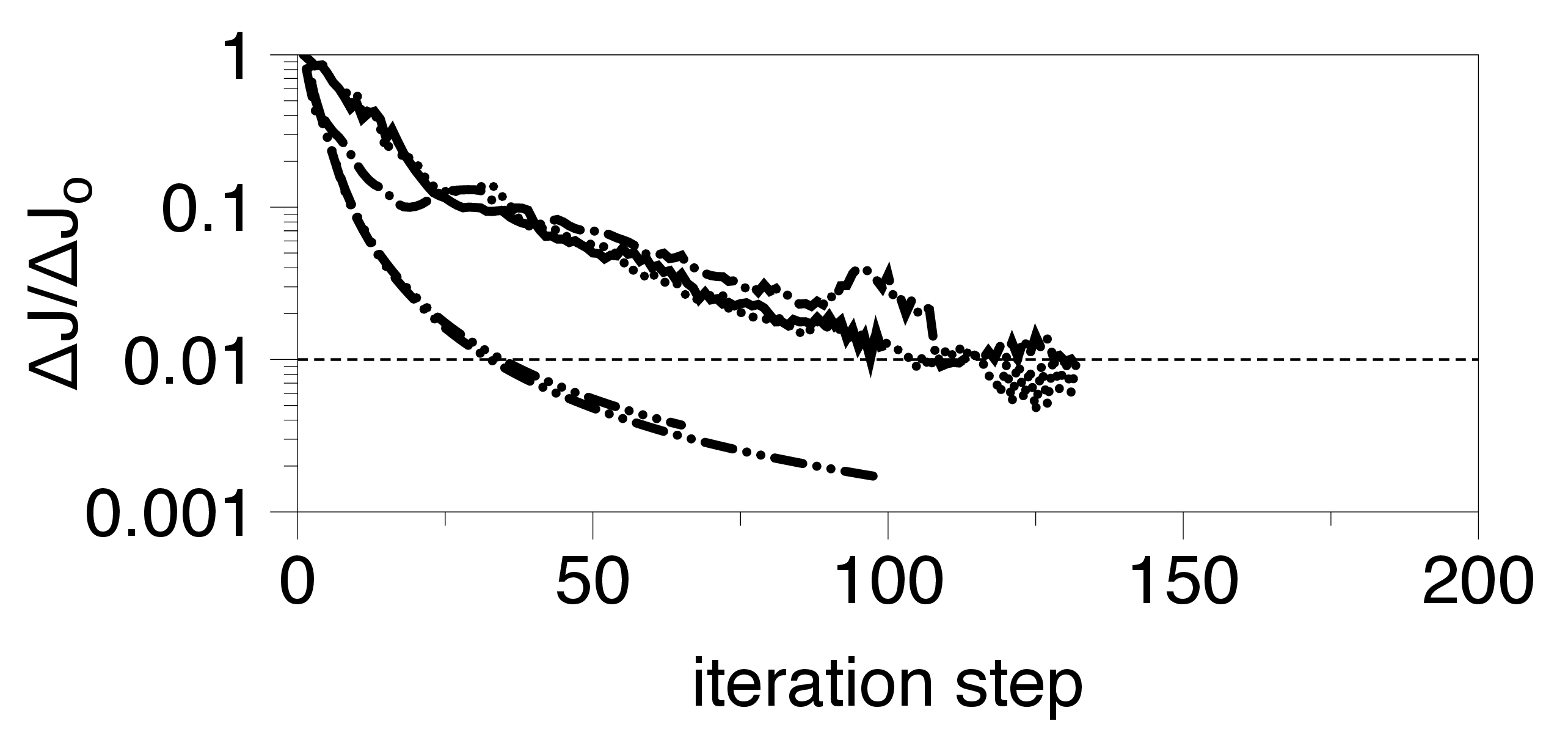}} \\
\caption{Evolution of cost-functional convergence rate
 starting from (a) circular and (b) random trajectories.
$(\solid) \; R_{21}=0.0$; \, $(\dotted) \; R_{21}=0.1$; 
\,$(\chndot) \; R_{21}=1.0$; \, $(\chndotdot)\; R_{21}=10.0$ \, $(\chndotdotdot)\; R_{21}\rightarrow \infty$. 
 A thin horizontal lines corresponds to $\Delta J / \Delta J_0$ which is used to judge 
 the convergence of each optimization.}
 \label{fig:convergence}
\end{figure}

First, we consider cases $A$ and $B$, in which the penalty of the sensor velocity is neglected. 
Figure~\ref{fig:convergence} shows the evolution of the cost functional convergence 
rate $\frac{ \Delta J }{ \Delta J_{o} }= \frac{ J_{1}^{n}-J_{1}^{n-1} }{  J_{1}^{1}-J_{1}^{0}  }  $ 
as a function of a number of iterations.
In the current study, the iteration for optimizing a sensor trajectory is continued
until the reduction of the cost functional becomes less than 1\% of the initial update, 
which is depicted by a thin horizontal line.
   
\begin{figure}
\flushleft
\subfloat[]
{\label{}
\includegraphics[width=0.5\textwidth]{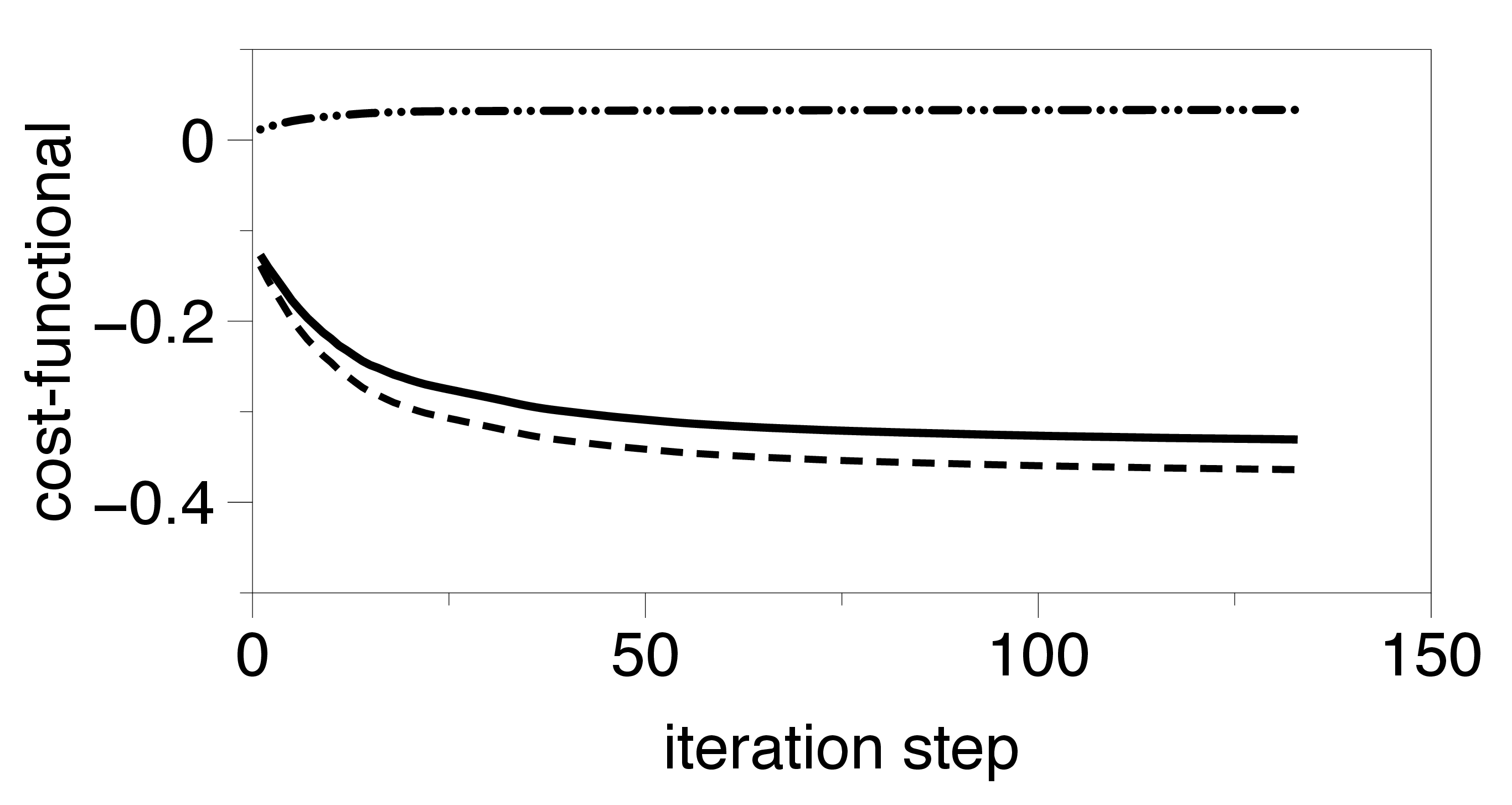}} 
\subfloat[ ]
{\label{}
\includegraphics[width=0.5\textwidth]{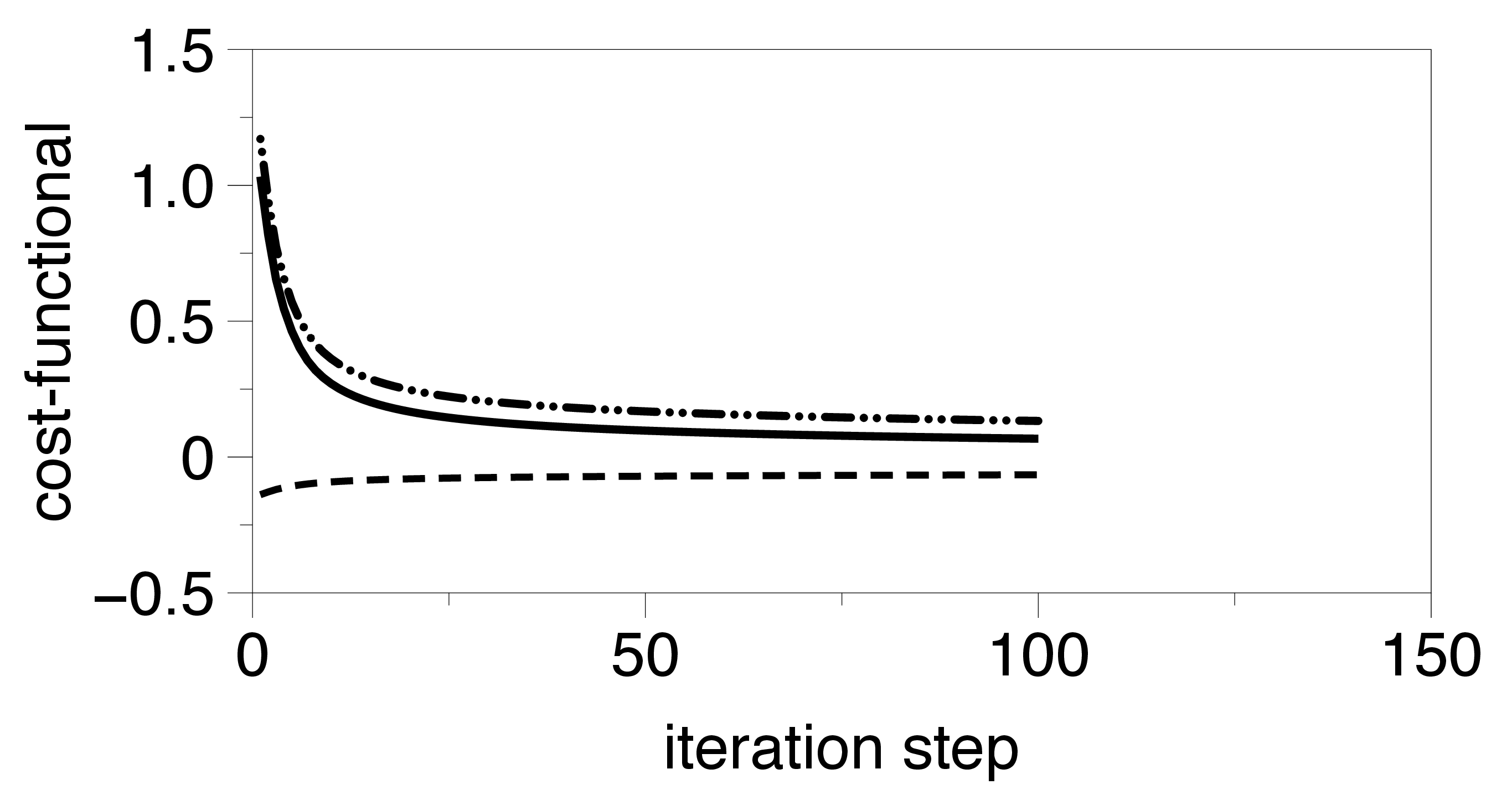}} \\
\caption{Evolution of the total cost-functional and  its
individual components for (a) $R_{21}=0.1$ and (b) $R_{21}=10$, for an initially random trajectory.
$(\solid)$: the total cost-functional,
$(\dashed)$: the first component,
$(\chndotdot)$: the second component. 
} 
 \label{fig:cost_func_01_10_a_b}
\end{figure}

Figure~\ref{fig:cost_func_01_10_a_b} shows the impact of $R_{21}$ on the evolution of the
cost functional~(\ref{cost_fun_or}) with the initial sensor trajectory generated by a random walk. 
Throughout the iterations, the first term in equation~(\ref{cost_fun_or}) is dominant for $R_{21} = 0.1$, 
whereas the second term becomes a primary factor for $R_{21} = 10$. These results indicate that 
the relative importance of the first and second terms in the cost functional~(\ref{cost_fun_or}) 
can be successfully controlled by changing $R_{21}$.
Note that similar trends are also confirmed when the optimization is initiated from a circular trajectory
(not shown here).

\begin{figure}
\centering
\includegraphics[width=0.75\textwidth]{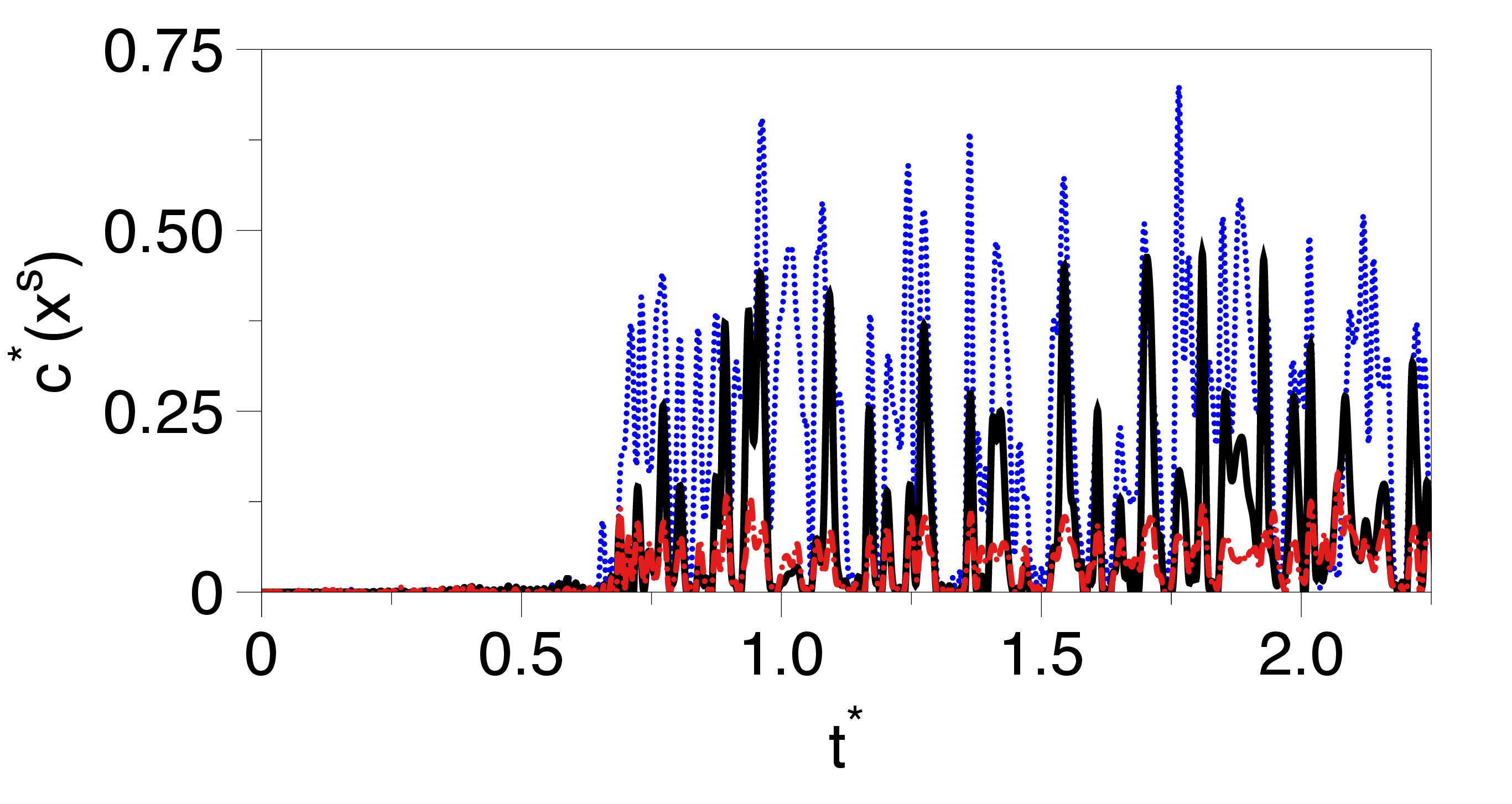} 
\caption{Time trace of the adjoint field at the source location for an initially random trajectory.
$(\solid)$: the initial trajectory
 { \color{blue} $\dotted$ }: $R_{21}=0.1$,
( {\color{red}$\chndot$}): $R_{21}=10$.
} 
 \label{fig:cstar_R01_1_10}
\end{figure}

Figure~\ref{fig:cstar_R01_1_10} shows the time trace of the adjoint field $c^*$ at the source 
location obtained by the sensor trajectories optimized with different $R_{21}$ from 
an initially random trajectory. Here, the horizontal axis is $t^{*} = T - t$, 
so that it proceeds backward in the original time $t$.
It can be observed that $c^{*}$ is absent during the initial period 
of $0 \leq t^{*} \leq 0.7$. This period corresponds to a time interval for 
the adjoint field generated at the sensor location reaches at the source location.
Indeed, considering the advection velocity at the channel center is $u_{c} \approx 18$,
and the streamwise distance between the sensor and the source is $L_{x}=12$, 
the convection time is approximated as $T_c \approx L_{x}/u_{c}  = 0.67$, 
which agrees well with the above initial period.

It is also found that the optimal sensor trajectories with large $R_{21}$ suppress 
the fluctuation of $c^*$, whereas those with small values tends to increase 
the intensity of $c^*$ with larger fluctuations. 
In order to quantify the fluctuation of $c^*$ relative to its mean, $\epsilon$ defined 
in equation~(\ref{epsilon_rms_mean}) is calculated and the results are listed for all cases in
Table~\ref{table:global_matrix}.
It is found that $\epsilon$ shows a non-monotonic relationship with $R_{21}$, and reaches
its minimum at $R_{21}=1$ regardless of the initial sensor trajectories. 
As will be shown later, large sensitivity with less fluctuation
in time is a key for better estimation, and $\epsilon$ is generally 
correlated with the resultant estimation performance.

\begin{figure}
\flushleft
\subfloat[]
{\label{fig:sensor_traj_yz_R01}
\includegraphics[width=0.5\textwidth]{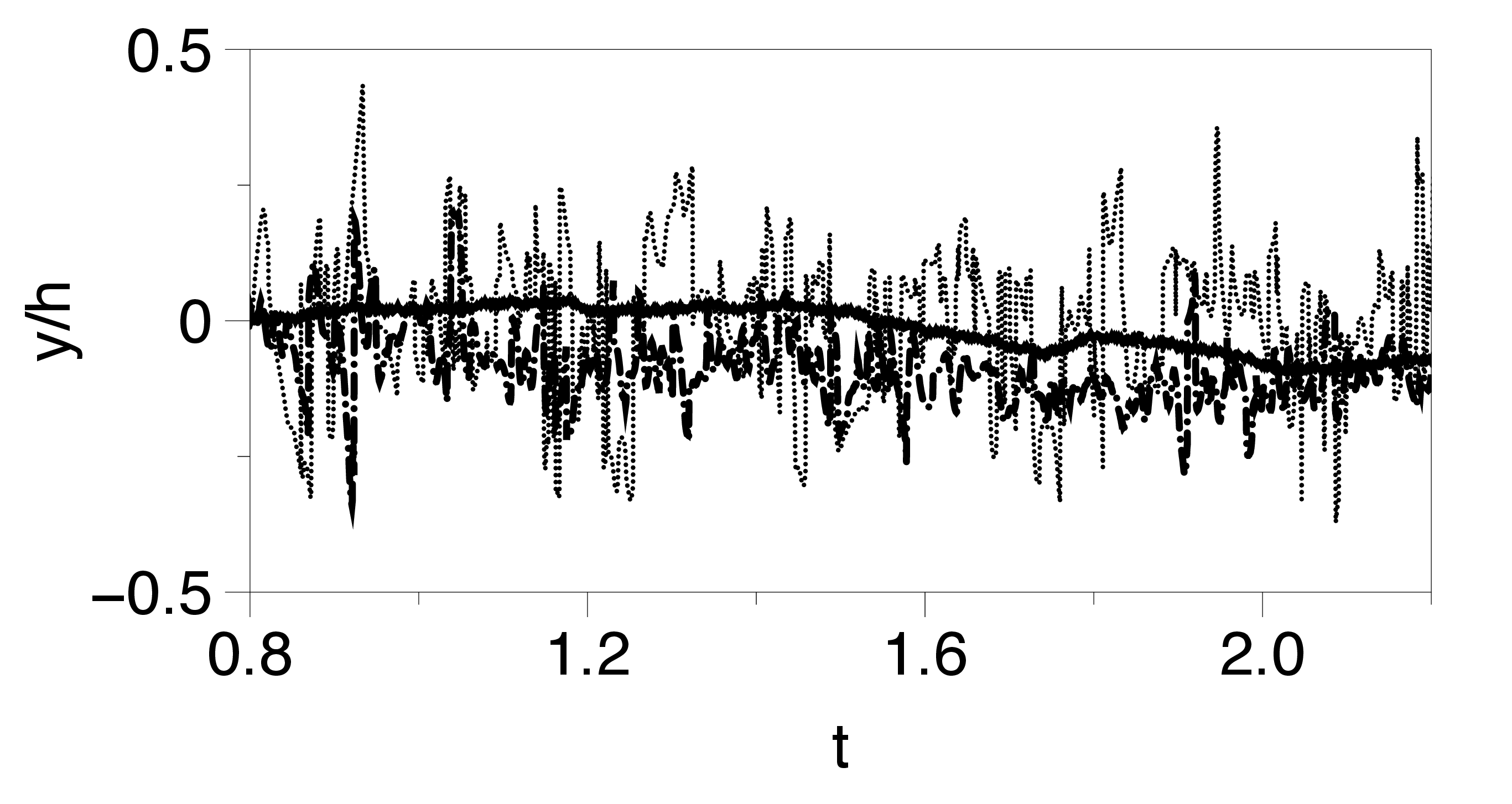}} 
\subfloat[ ]
{\label{fig:sensor_traj_yz_R10}
\includegraphics[width=0.5\textwidth]{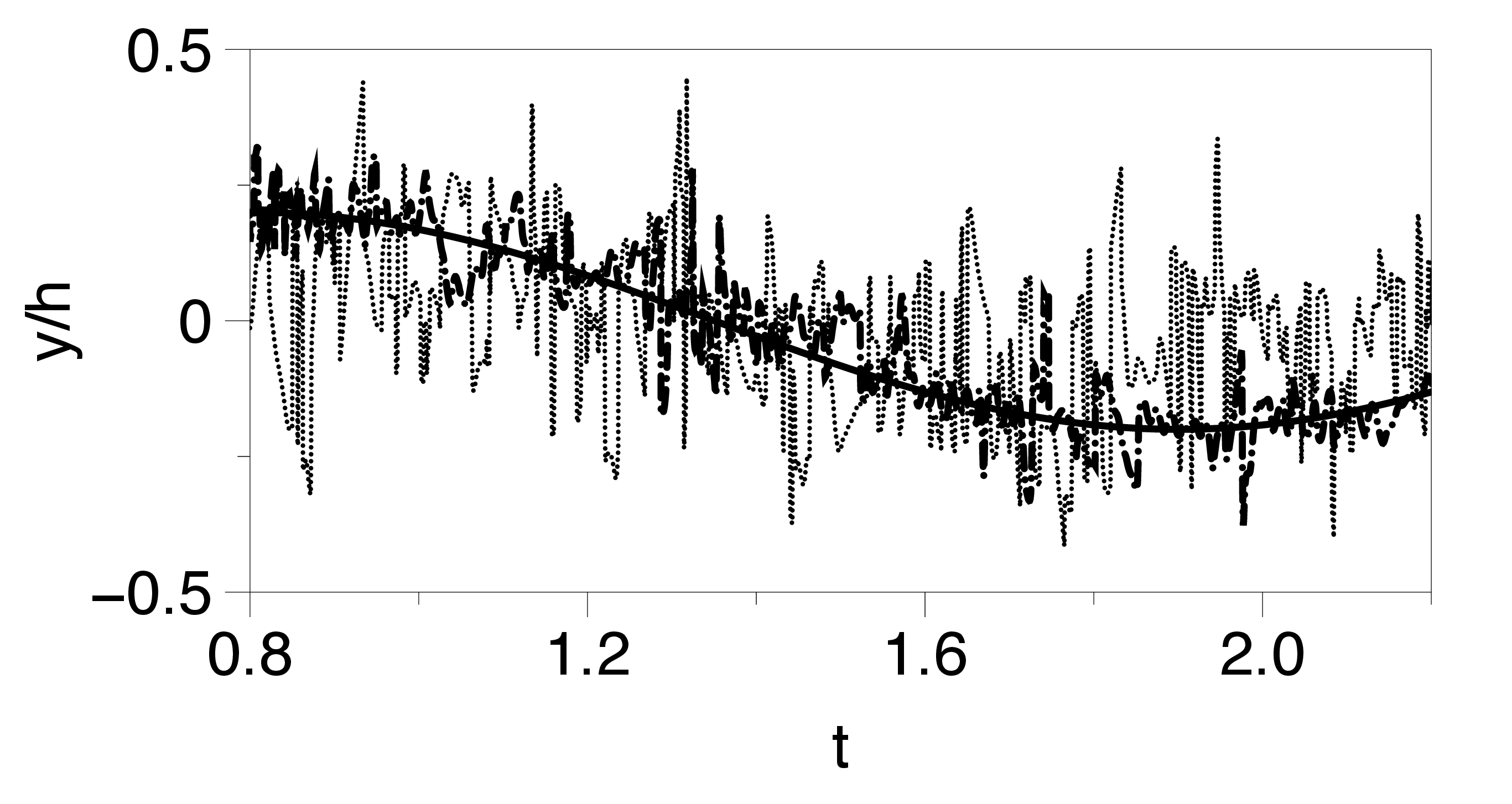}} 
\caption{Time trace of the sensor's trajectory along the $y$ coordinate starting from (a) random and 
(b) circular trajectories.
($\solid$): the initial trajectory, 
($\dotted$): the optimal trajectory with $R_{21}=0.1$,
($\chndot$): the optimal trajectory with $R_{21}=10$.
} 
 \label{fig:sensor_traj_y_t_init_r01_10}
\end{figure}

Figure~\ref{fig:sensor_traj_y_t_init_r01_10} shows the time trace of the $y$-coordinate of the 
sensor trajectories optimized from random and circular ones.

It is found that a smaller $R_{21}$ yields more spread sensor trajectory from the center line, 
i.e., $y = 0$, and quicker changes of sensor location, indicating 
high sensor velocity and abrupt changes in its sign.
Furthermore, the correlation coefficient between the optimal trajectories starting 
from circular and random trajectories is found to increase with decreasing value of $R_{21}$. 
The above observations could be explained by the source term appearing in the last term on the RHS of  
equation~\eqref{theta_evol_eq}. For small $R_{21}$, this term tends to become constant, thus 
independent of sensor trajectory. 

A similar trend is also observed for $z$ coordinate 
of the sensor trajectory (not shown here).
\\

\subsubsection{Case $C$}
\begin{figure}
\flushleft
\subfloat[]
{\label{fig:sensor_traj_pdf_R01}
\includegraphics[width=0.5\textwidth]{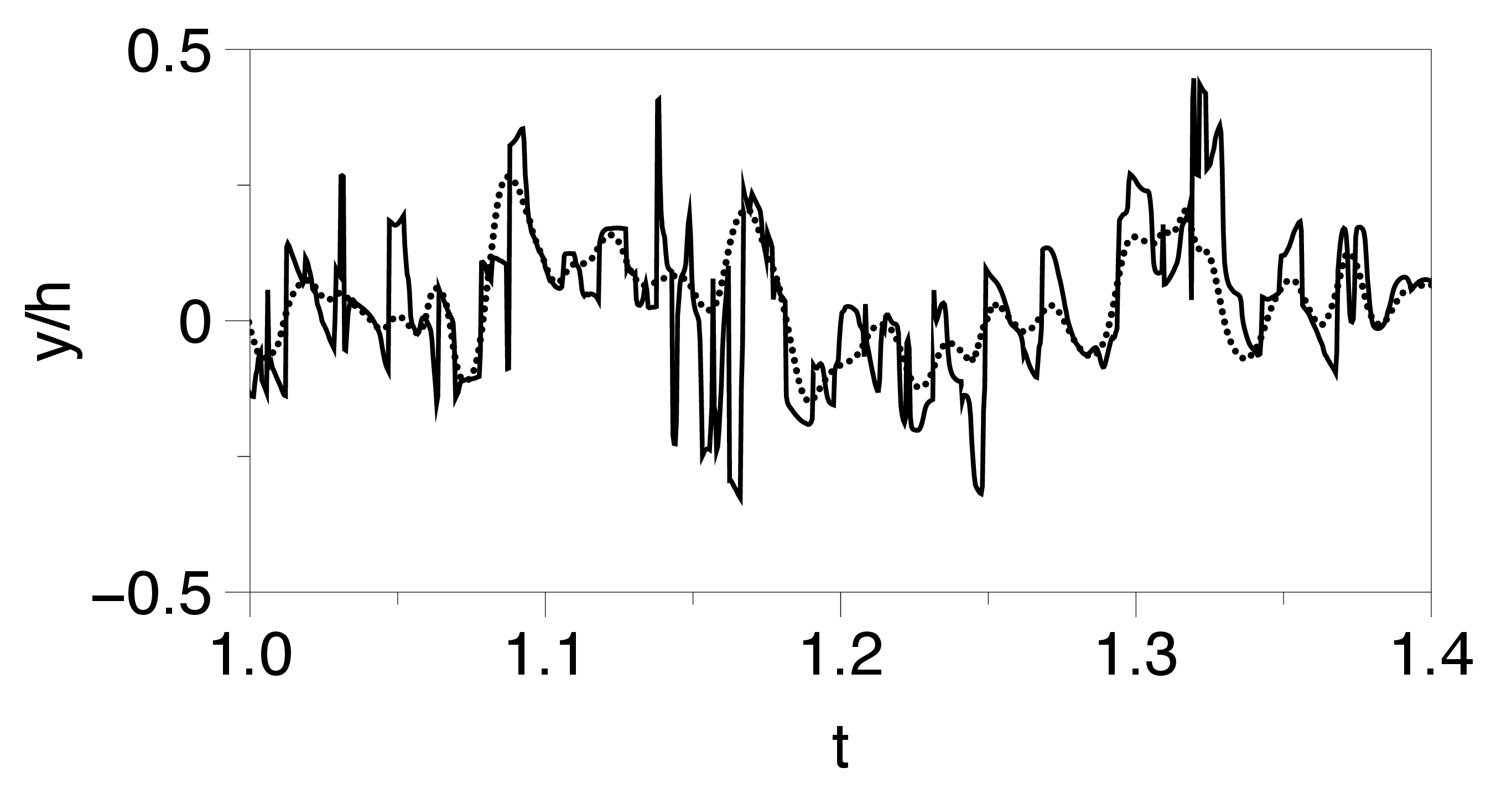}
} 
\subfloat[ ]
{\label{fig:sensor_traj_pdf_R10}
\includegraphics[width=0.5\textwidth]{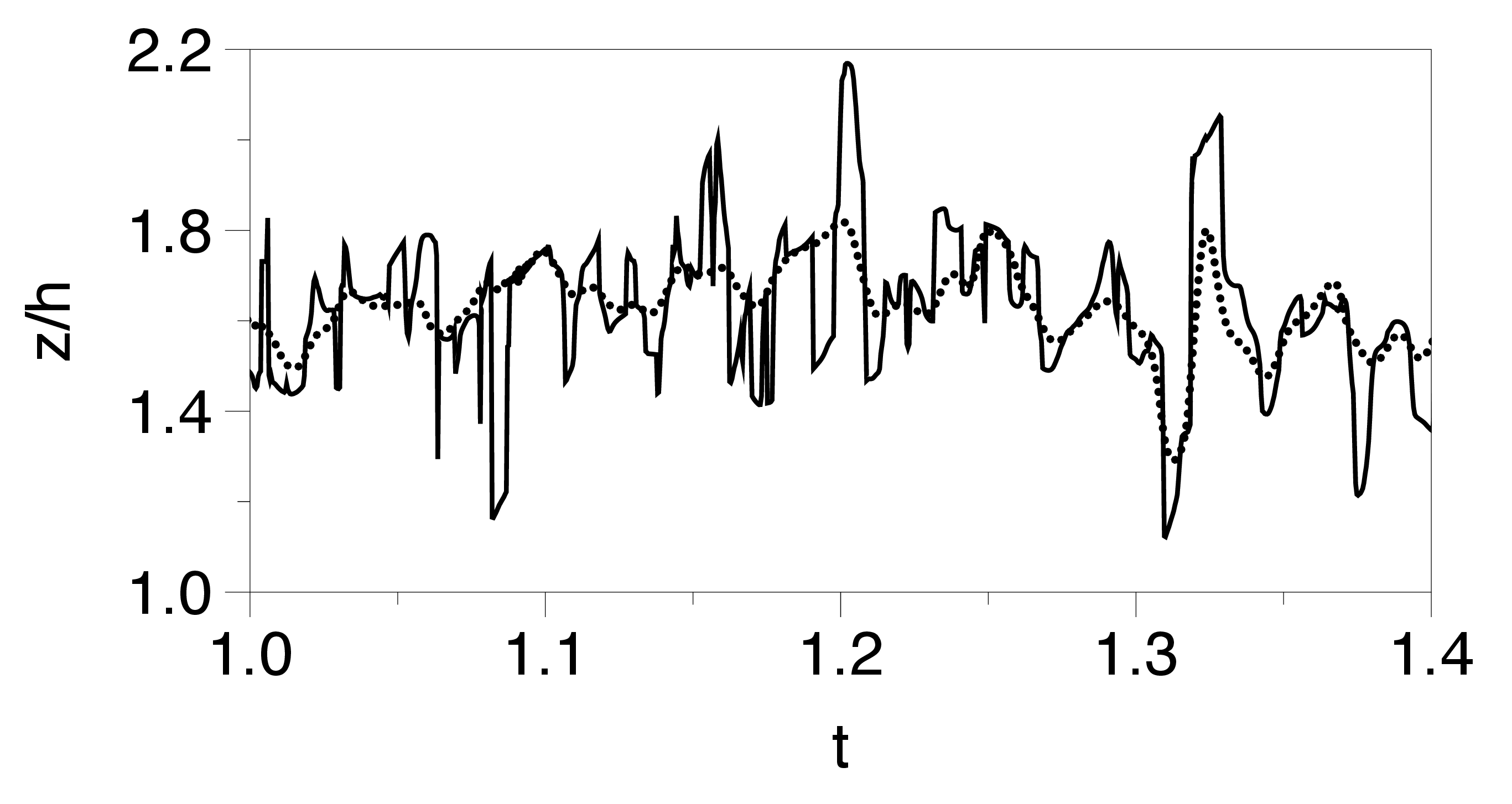}
}
\caption{ Time traces of the (a) normal  and (b) spanwise coordinate 
of the optimal sensor trajectory starting from a random trajectory. 
$(\solid)$: Case B2
$(\dotted)$: Case C
} 
 \label{fig:yz_over_t_RA3_0_1_RAND_nolabel_onlyline}
\end{figure}

Figure~\ref{fig:yz_over_t_RA3_0_1_RAND_nolabel_onlyline} shows the time traces of 
the sensor trajectories with and without the penalty of the sensor speed for Cases $C$ and $B2$.
It can be seen that non-zero value of $R_{31}$ in Case $C$ yields a smoother trajectory
 compared to $R_{31}=0$ (Case $B2$). This is consistent with the presence of the penalty term,
i.e., the third term in the cost functional~(\ref{cost_fun_or}).
  
\begin{figure}
\flushleft
\subfloat[]
{\label{fig:uy_over_ubulk}
\includegraphics[width=0.5\textwidth]{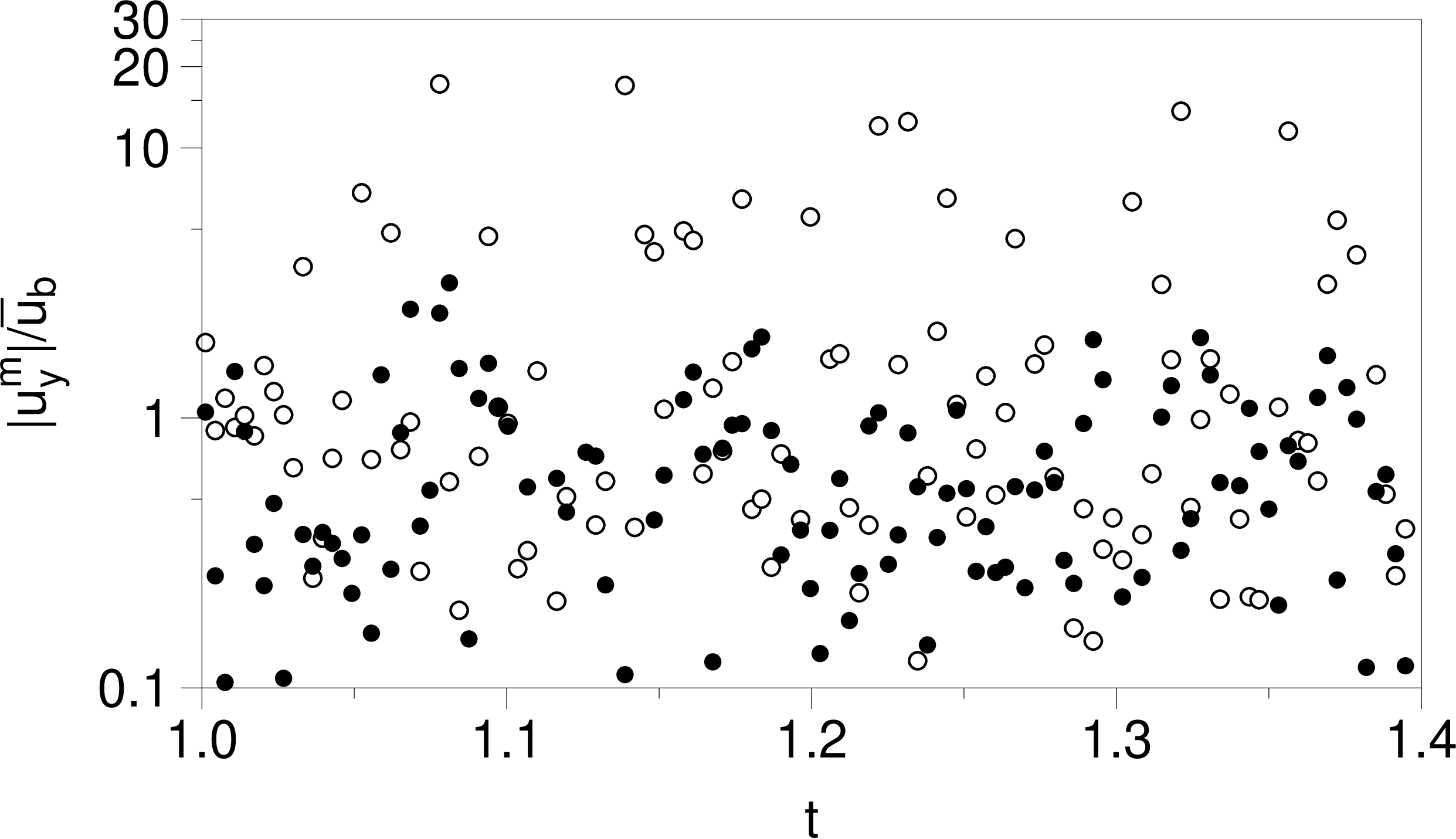}
} 
\subfloat[ ]
{\label{fig:uz_over_ubulk}
\includegraphics[width=0.5\textwidth]{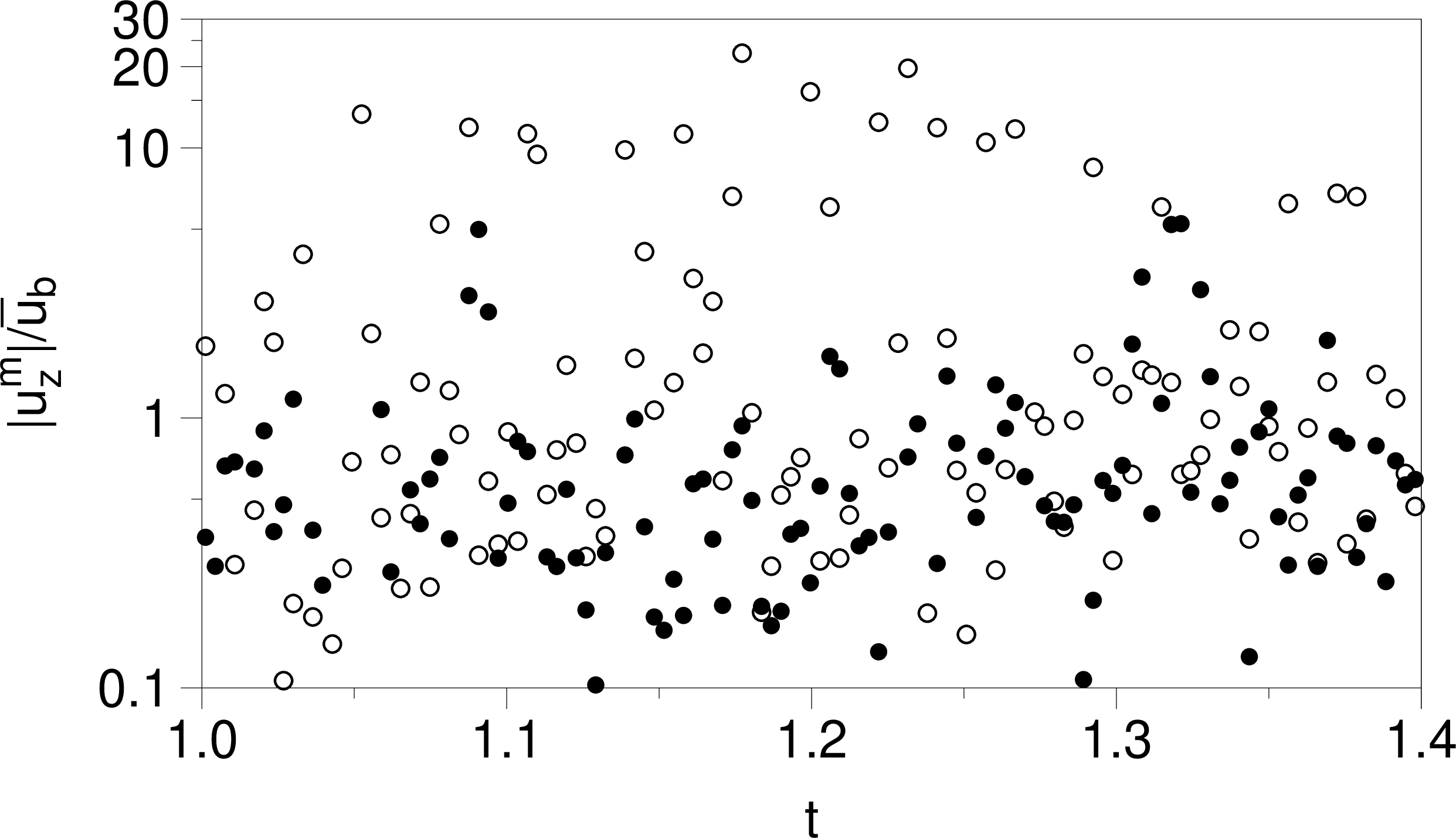} 
} 
\caption{ \textit{Top:} Time traces of sensor velocity along (a) wall-normal and (b) spanwise
directions, normalized by the bulk mean velocity.   
Open and filled circles correspond to Cases $B2$ and $C$, respectively.
} 
\label{fig:velocities_over_fluid_velocity} 
\end{figure}

Figures~\ref{fig:velocities_over_fluid_velocity}(a)-(b) show the time traces of the sensor speed 
for the two cases in the $y$ and $z$ directions, respectively. 
Here, the sensor speed is normalized by the bulk mean velocity $\bar{u}_{b}$. 
It is found that, for Case B2 where no penalization on the sensor velocity is imposed,
the sensor velocity intermittently rises beyond the bulk mean velocity as shown by open circles in Figs.~\ref{fig:velocities_over_fluid_velocity}(a)-(b).
In contrast, by introducing the penalization of the sensor velocity in Case $C$, 
the abrupt increase of the sensor velocity is suppressed, so that 
the sensor velocity is mostly below the bulk mean velocity.

\begin{figure}
\centering
\includegraphics[width=0.6\textwidth]{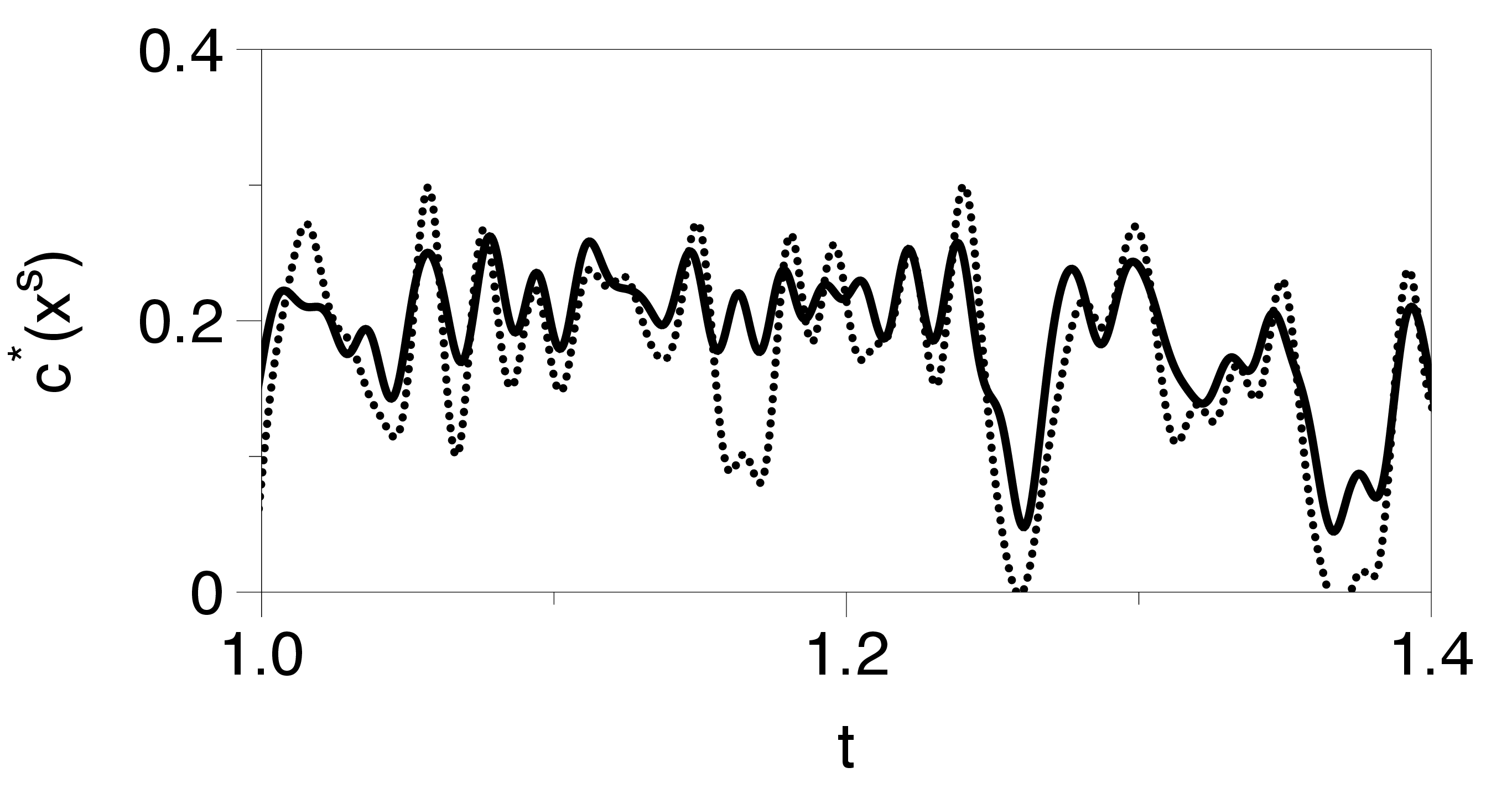} 
\caption{Time trace of the adjoint field at the source location generated from
the trajectories optimized from the initially random trajectory. 
$\solid$: Case B2, 
$\dotted$: Case C
} 
 \label{fig:cstar_RA3_01_10}
\end{figure}

Figure~\ref{fig:cstar_RA3_01_10}  shows the time traces of the adjoint fields at the source location 
which are generated from the optimal trajectories in Cases $B2$ and $C$. 
It is observed that the fluctuation of the adjoint field becomes larger in Case $C$, where
the penalty of the sensor speed is introduced. This implies that the smoother sensor trajectory
is obtained by compromising the stabilization of the temporal fluctuation of the adjoint field. 
This is also reflected to the increase of $\epsilon$ in Case $C$ (see, Table~\ref{table:global_matrix}).

\subsection{Scalar source estimation with optimal sensor trajectory}
In the previous subsection, it was shown that the parameters $R_{21}$ and $R_{31}$ defined by
equations~(\ref{R_ratio}, \ref{R_ratio2}) successfully change the relative importance of the three terms 
comprizing the cost functional~(\ref{cost_fun_or}), and the corresponding optimal sensor trajectories
are obtained under the compromise of the three terms. Here, we evaluate the performances of 
scalar source reconstruction based on the signals obtained from the optimal trajectories.

Table~\ref{table:global_matrix} summarizes the reconstruction performances
evaluated by the correlation coefficient $\psi^\phi$ and the L2 norm $\ell^{2}$.
It is found that the optimal trajectories generally perform much better than 
the initial circular/random trajectory, and also the single stationary sensor.
It should be also noted that the highest performance is commonly
obtained in Case $B2$ regardless of the initial sensor trajectory.

\begin{figure}
\subfloat[B1]
{
\includegraphics[width=0.45\textwidth]{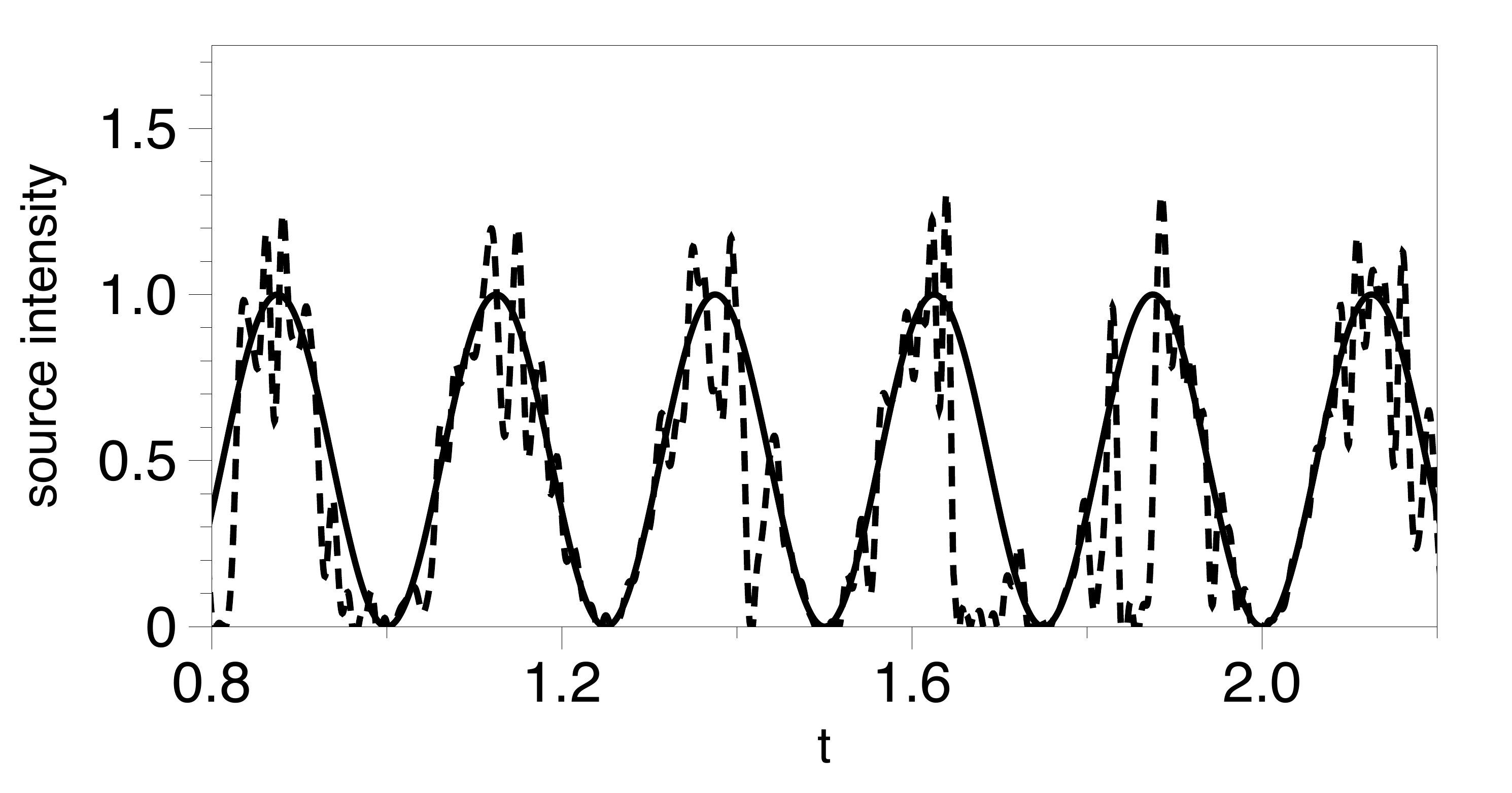}
} 
\subfloat[B2]
{
\includegraphics[width=0.45\textwidth]{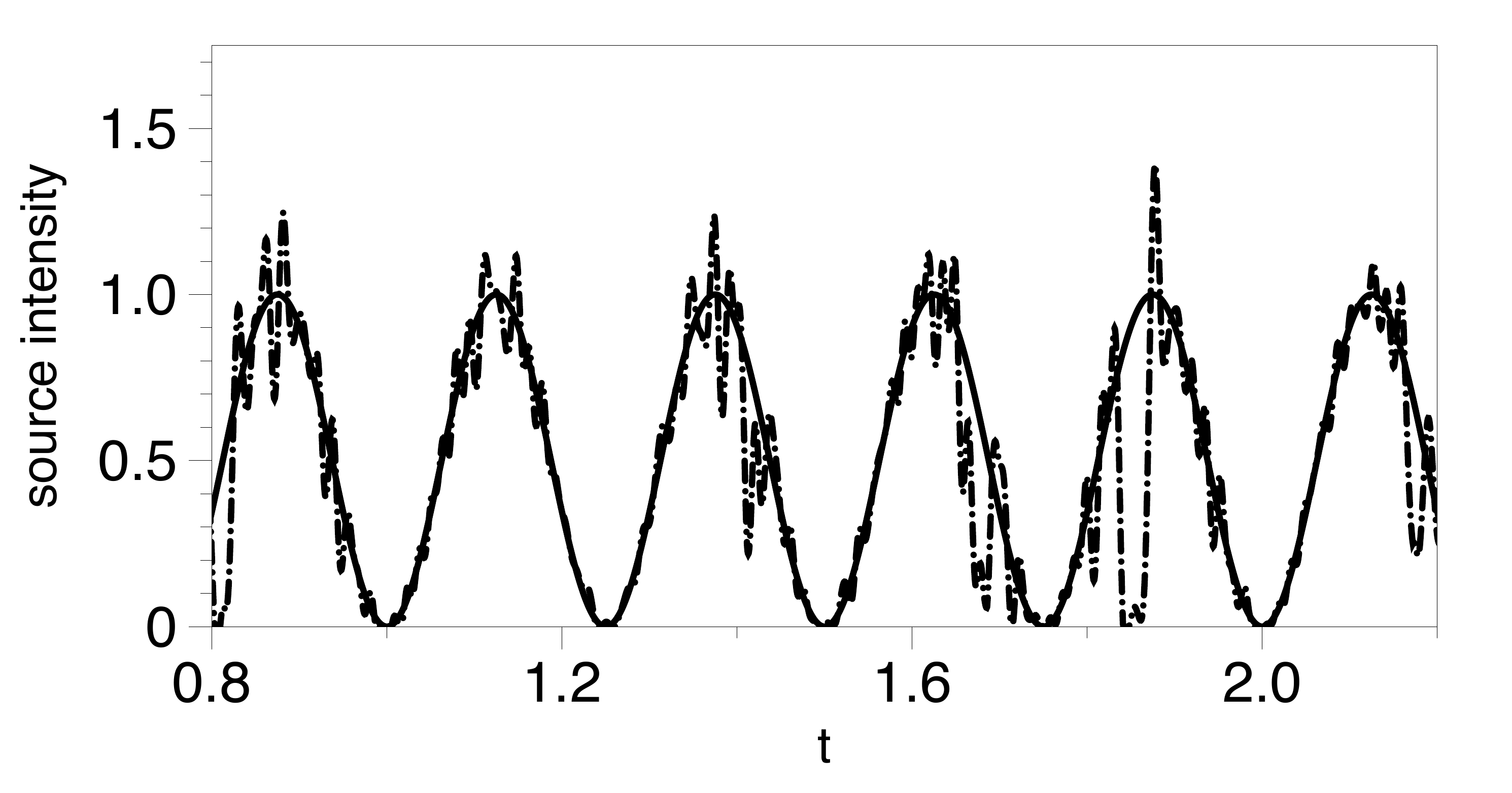}
} \\
\subfloat[B3]
{
\includegraphics[width=0.45\textwidth]{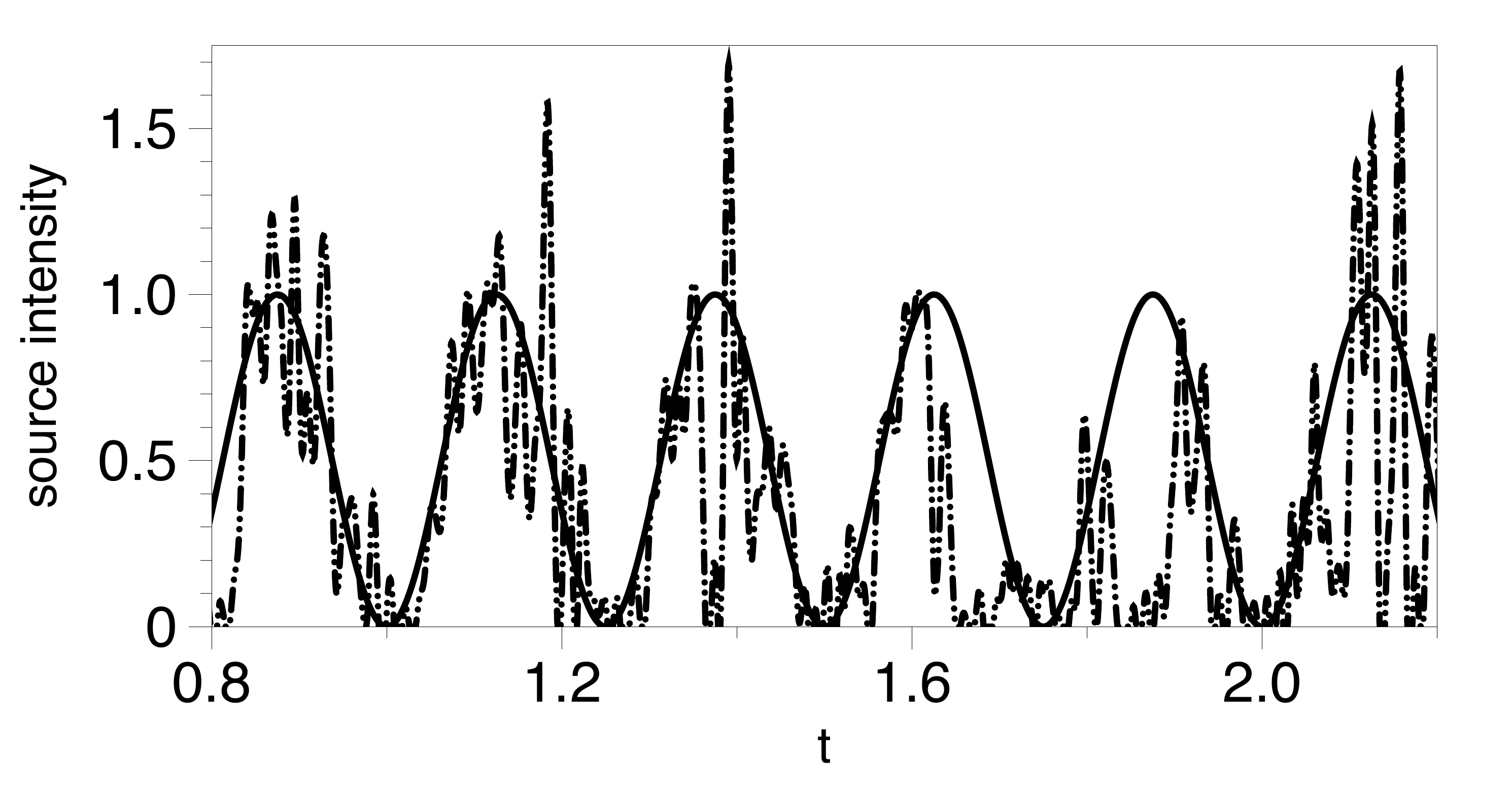}
} 
\caption{Time traces of the source intensity estimated 
with trajectories optimized from the initial random trajectory. 
Solid line ($\solid$): true profile,  dotted line ($\dotted$): Case $B1$,
dashed dotted line ($\chndot$): Case $B2$, 
dashed double dotted line ($\chndotdot$ ): Case $B3$.
Only a limited time frame of the entire time horizon considered is shown. }
\label{fig:phi_rand_01_1_10}
\end{figure}

\begin{figure}
\centering
\includegraphics[trim=2cm 3cm 4cm 3cm, clip=true, totalheight=0.4\textheight]{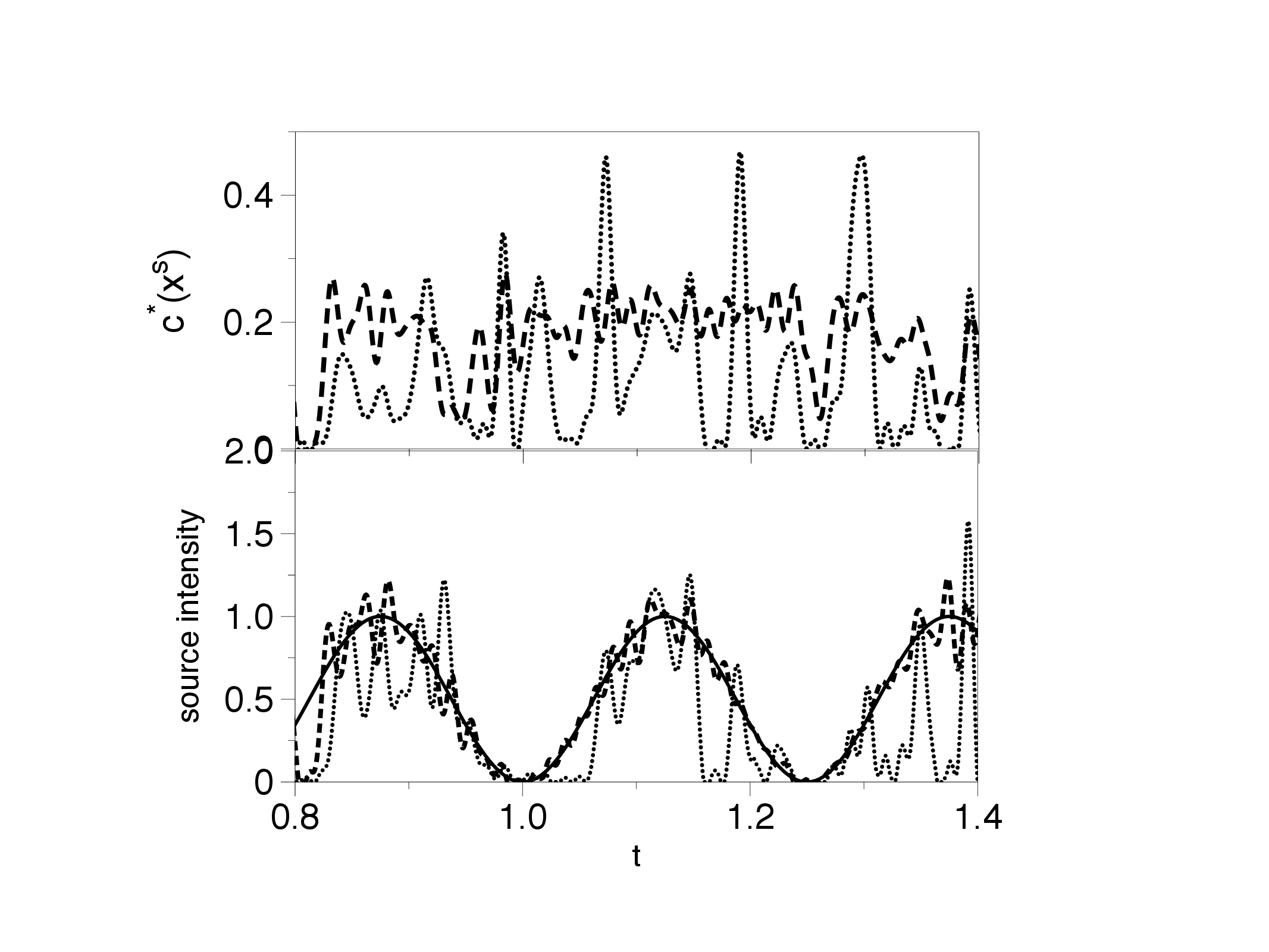} 
\caption{
Comparison of the time trances of (Top) the adjoint field $c^*$ at the source location, and (Bottom)
the resultant estimation of scalar source intensity for the initial random trajectory and
the optimal trajectory obtained in Case $B2$. Dotted line ($\dotted$): initial random trajectory,
Dashed line ($\dashed$): optimal trajectory, Solid line ($\solid$): true source intensity.
} 
 \label{fig:theoretical_explanation}
\end{figure}

In figure~\ref{fig:phi_rand_01_1_10}, the true and estimated scalar source intensities are plotted.
Consistent with Table~\ref{table:global_matrix}, the estimation with the optimal trajectory 
in Case $B2$ shows the best match with the true profile. 
In order to explain the best estimation performance in Case $B2$, 
the adjoint scalar field at the source location for the initial random trajectory and 
the optimal trajectory are compared in the top figure of figure~\ref{fig:theoretical_explanation},
whereas the resultant reconstruction of the scalar source is shown in the bottom.
It can be seen that the estimation error is strongly correlated with the fluctuation of the adjoint field 
at the source location. A larger fluctuation generally causes a large deviation of the estimation 
from the true profile. Especially, the presence of time periods with no significant adjoint field 
in the initial trajectory results in the failure of estimation during these periods.
These results justify the current setting of the cost functional~(\ref{cost_fun_or}) and 
also suggest that $\epsilon$ can be used as a diagnostic parameter 
for the estimation capability. Indeed, as shown in Table~\ref{table:global_matrix}, 
the estimation performance is negatively correlated with $\epsilon$.

We also address the effects of the penalty of the sensor speed on 
the estimation performance. As shown in figure~\ref{fig:cstar_RA3_01_10},
the adjoint field at the source location in Case $C$ shows larger fluctuation
than that in Case $B2$. As a result, $\epsilon$ is increased and the estimation
performance is deteriorated (see, Table~\ref{table:global_matrix}). 
Adding the penalty term in the cost functional~(\ref{cost_fun_or}) results in 
a smoother trajectory at the cost of larger fluctuation of the adjoint field 
at the source location. This explains the worse performance in Case $C$.

\subsection{Cost functional with $\epsilon$}

One of the main issues in the cost functional~(\ref{cost_fun_or}) is that
it requires to find the optimal values of the weighting coefficients, i.e., $R_{21}$ and $R_{31}$. 
Especially, the value of $R_{21}$ is important to determine the relative importance of 
the first and second terms in equation~(\ref{cost_fun_or}). According to the present results, however, 
the estimation performance is well correlated with a single quantity,
i.e., $\epsilon$, throughout all the cases considered.
This motivates us to define the following cost functional:
\begin{equation}\label{alter_costfunc}
J_{2} = \epsilon = \frac{ c^{*}_{rms}(\bold{x}^{s}) }{ \overline{c^{*}}(\bold{x}^{s}) }\,.
\end{equation}
The advantage of introducing the above cost functional is that there is no adjustable parameter.

Following the same procedure as described in $\S$~\ref{optimization_strategy},
we obtain the following adjoint-of-the-adjoint equation for $\theta$ (see Appendix~B for the detailed derivation):
\begin{equation}\label{theta_alt_eq}
\frac{\partial \theta}{\partial t} + \frac{\partial (\theta u_{j})}{\partial x_{j}}
=\bigg\{ \frac{\partial }{\partial x_{j} }\left(\frac{1}{Pe}\frac{\partial \theta }{\partial x_{j} }\right)\bigg\}
+  \frac{1}{T}\left( \frac{  c^{*}_{rms}  }{\overline{c^{*}}^{2}  }-
\frac{ [c^{*}-\overline{c^{*} }  ] }{ \overline{c^{*} } c^{*}_{rms}  }
 \right)\delta ( \bold{x}-\bold{x}^{s} )\,,
\end{equation}
whereas the formula for updating the sensor trajectory is the same as equation~(\ref{update_exp}). 
The above equation can be regarded as a modified version of equation~\eqref{theta_evol_eq},
in which the weighted coefficients $\alpha_{1}$ and $\alpha_{2}$ are replaced with
the statistics of $c^*$.
As for the initial condition of the sensor trajectory, 
we consider a stationary sensor located directly downstream of the source. 
This location was chosen, since it results in the minimum $\epsilon$ in the sensing plane 
when a sensor is stationary. 

\begin{figure}
\flushleft
\subfloat[]
{\label{fig:cstar_tstar_center}
\includegraphics[width=0.5\textwidth]{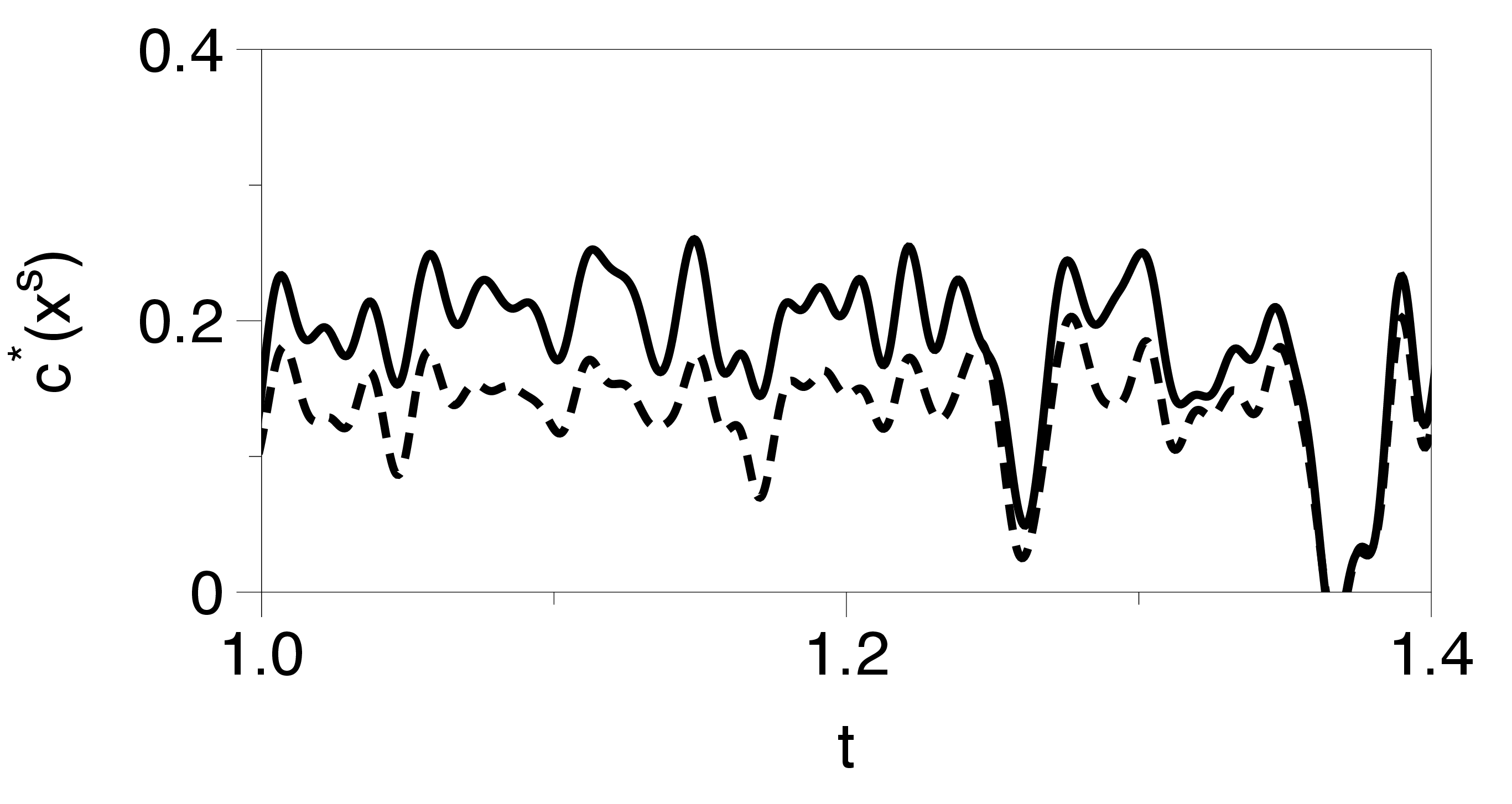}
} 
\subfloat[ ]
{\label{fig:y_t_cent_JMOD_R1}
\includegraphics[width=0.5\textwidth]{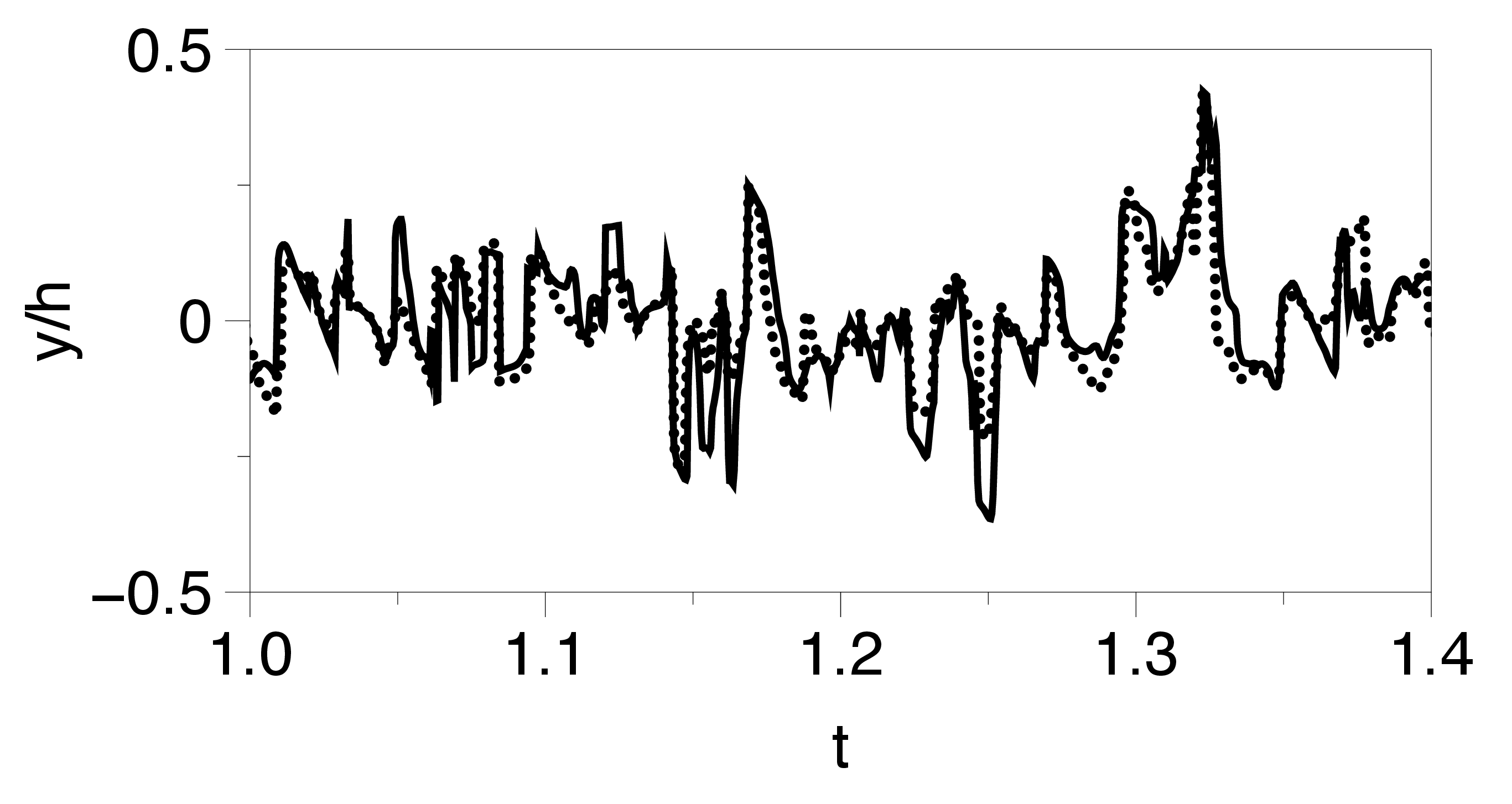}
} 
\caption{ Comparison between the alternative strategy 
($\dashed$) 
and the optimal case 
$R_{21}=1$ of the initial strategy 
($\solid$), 
for a sensor initially at rest at the centerline point. 
(a) Time trace of the adjoint field at the source location for a limited time horizon. 
(b) Time trace of the normal coordinate of sensor trajectory.} 
 \label{fig:cstar_pdf_center_jmod_r1}
\end{figure}

Figure~\ref{fig:cstar_pdf_center_jmod_r1} (a) shows the adjoint fields at the source location
for the sensor trajectories optimized by the original cost functional~(\ref{cost_fun_or})
with $(R_{21}, R_{31}) = (1.0, 0)$ (Case $B2$) and the newly introduced cost functional~(\ref{alter_costfunc}).
Although the new cost functional yields a slightly lower adjoint field 
than the original one, the values of $\epsilon$ in both cases
are similar and around $0.85$ after optimization.

The time traces of the wall-normal coordinate of the optimal sensor trajectories are
compared in figure~\ref{fig:cstar_pdf_center_jmod_r1} (b). It is confirmed that the sensor 
trajectories obtained from the original and new cost functionals are similar.
The results shown in figure~\ref{fig:cstar_pdf_center_jmod_r1}  indicate that 
the new cost functional~(\ref{alter_costfunc}) without any adjustable parameter 
is a potential alternative to the original cost functional~(\ref{cost_fun_or}) 
with the optimal values of $R_{21}$ and $R_{31}$.

\begin{figure}
\centering
\includegraphics[width=0.7\textwidth]{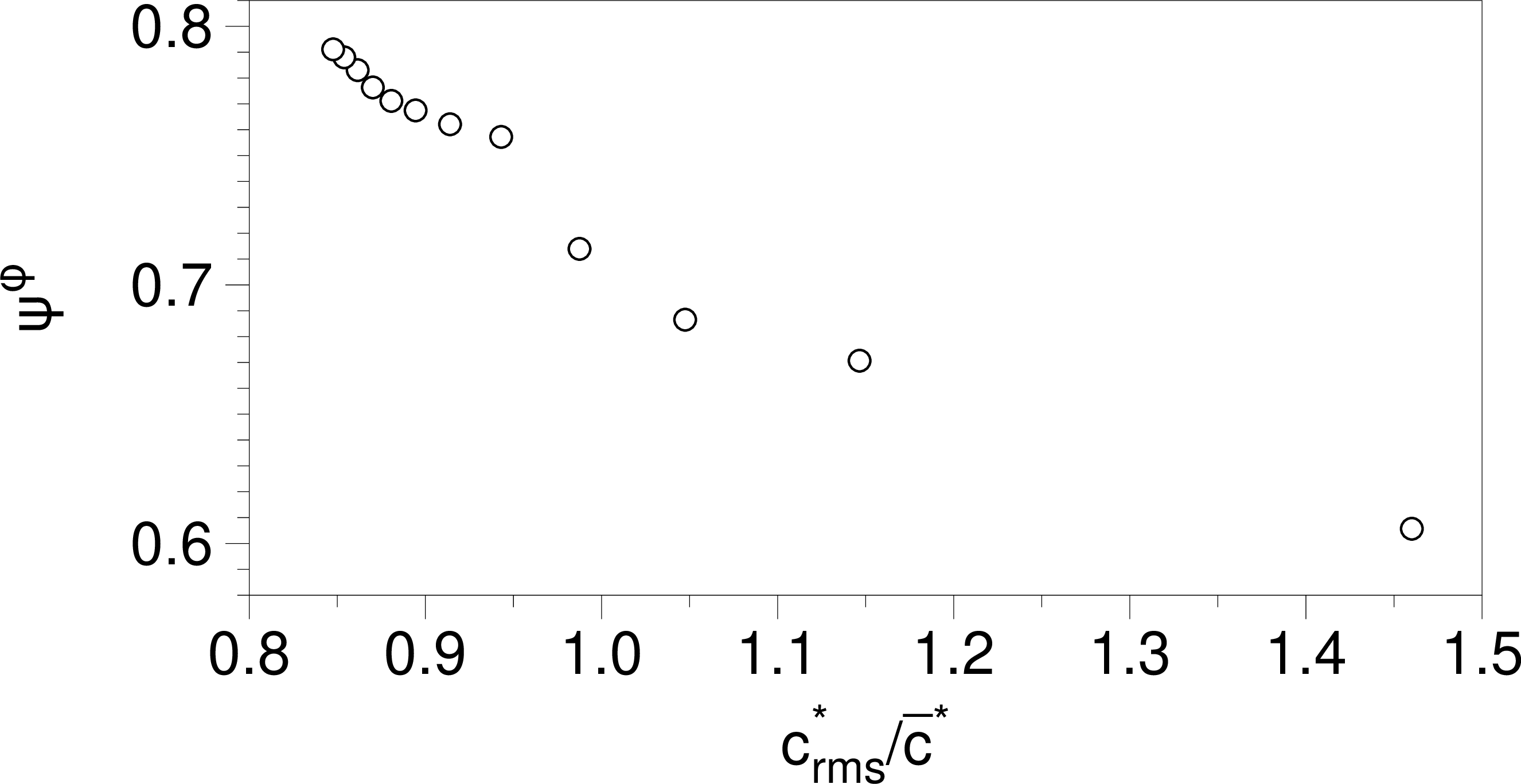} 
\caption{ Correlation coefficient $\psi^{\phi}$ as a function of adjoint ratio 
$c^{*}_{rms}/\bar{c}^{*}$ for a a mobile sensor whose trajectory is optimized with the new 
cost functional $J_{2}$. Each open circle refers to a different iteration step of the iterative 
process. }
 \label{fig:sensor_performance}
\end{figure}

In order to further validate the effectiveness of the new cost functional $J_{2}$,
Fig.~\ref{fig:sensor_performance} shows the evolution of the estimation performance, i.e., the correlation coefficient $\psi^{\phi}$
between the true and estimated source profiles, as a function of $J_{2}$ 
obtained at every ten iterations in the optimization process.
It can be confirmed that the estimation performance is increased with decreasing $J_{2}$.

\begin{figure}
\centering
\includegraphics[width=0.7\textwidth]{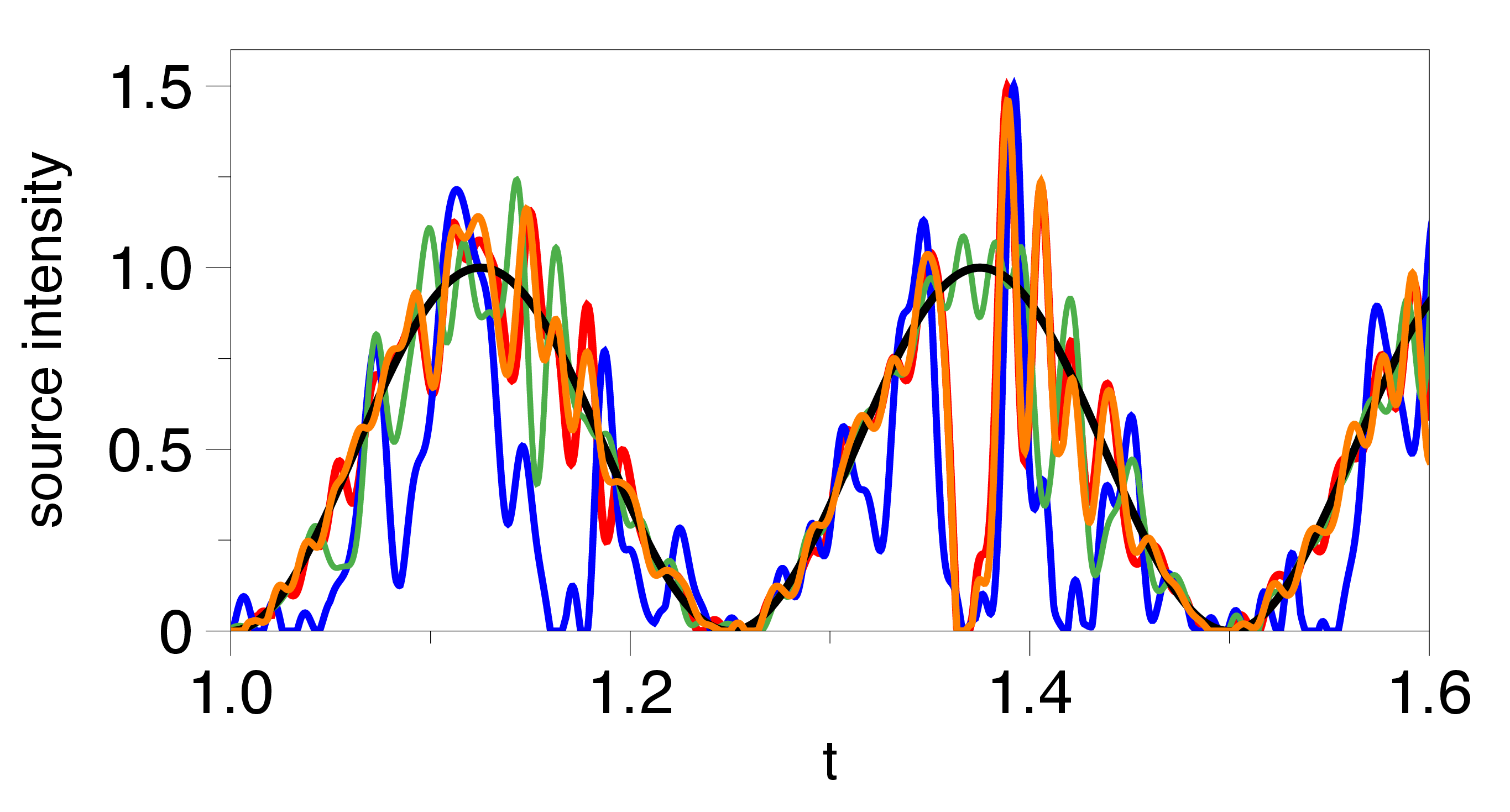} 
\caption{ Time evolution of the reconstructed source profiles for $f=4$.
black: true profile, 
red: a mobile sensor whose trajectory is optimized with the new cost functional~(\ref{alter_costfunc}),
orange: a mobile sensor whose trajectory is optimized with the original cost-functional (case $B2$),  
blue: a single stationary sensor,
green: $17$ stationary sensors.
Only a limited time frame of the entire time horizon is shown. }
 \label{fig:phi_frequency_4_mov_fix_mult}
\end{figure}

The time traces of the true and reconstructed source intensity at $f=4$
are shown in figure~\ref{fig:phi_frequency_4_mov_fix_mult}. 
The figure includes the estimated source intensities based on a single mobile sensor whose trajectory
is optimized by the cost functional $J_1$ in Case $B_2$ and the new cost functional $J_2$.
Also, the estimations with a single and seventeen stationary sensors are plotted for comparison.
It can be confirmed that the estimation performance of the single sensor moving along the optimal trajectory
obtained by the new cost functional $J_2$ is as good as that of seventeen stationary sensors.

The present optimization strategy is based on the idea to maintain the sensor sensitivity 
at a source location throughout the entire time horizon. Hence, the resultant optimal 
trajectory should be effective at different pulsating frequencies. 
Table~\ref{table:frequency_all_results} summarizes the estimation quality obtained 
from a single movable sensor whose trajectory is optimized by the new cost functional~(\ref{alter_costfunc}), 
and single and multiple stationary sensors at different pulsating frequencies of 
the source, i.e., $f = 2, 4, 8$ and $16$.
We observe that the estimation performance gradually deteriorates with increasing $f$
in all cases. However, the single mobile sensor with the optimal trajectory shows significant improvement
compared with the single stationary sensor, and its performance is close to that of seventeen stationary sensors.

\begin{table}
\centering
\begin{tabularx}{\textwidth}{@{}lYYYYYYY@{}}
\multirow{3}{*}{$f$} &  \multicolumn{4}{  c }{Stationary}                       & \multicolumn{2}{ c }{\multirow{2}{*}{Single Mobile ($J_{2}$)}}              \\ 
                     &  \multicolumn{2}{c}{\multirow{1}{*}{Single}}              & \multicolumn{2}{ c }{\multirow{1}{*}{Multiple ($n=17$) }}   & \multicolumn{2}{ c }{\multirow{2}{*}{}}  \\ 
                     & $\psi^{\phi}$     & $\ell^{2}_{norm}$  & $\psi^{\phi}$    & $\ell^{2}_{norm}$ &    $\psi^{\phi}$ & $\ell^{2}_{norm}$  \\ 
    2                &    0.54           &    0.40            & 0.96             &             0.10  &       0.83       &    0.22  \\ 
    4                &    0.60           &    0.38            & 0.94             &             0.13  &       0.80       &    0.24 \\ 
    8                &    0.49           &    0.41            & 0.93             &             0.13  &       0.84       &    0.21  \\ 
    16               &    0.49           &    0.40            & 0.93             &             0.13  &       0.86       &    0.20 \\ 
\end{tabularx}
\caption{Comparison of the estimation performances at different source frequencies obtained by
a single and seventeen stationary sensors and a single mobile sensor whose trajectory is optimized 
under the new cost-functional $J_{2}$.}
\label{table:frequency_all_results}
\end{table}


\section{Summary and Conclusions}

In the present study, we consider a problem of scalar source estimation in a turbulent environment.
Assuming that the location of a point scalar source is known, the time dependence of the source intensity 
is estimated based on the signals obtained from a sensor located downstream. Particularly, we focus on
developing a new strategy for optimizing a trajectory of a moving sensor.

The key idea behind the present optimization strategy is to maximize the sensor 
sensitivity and minimize its temporal fluctuation at the source location, 
whereas the sensor sensitivity is theoretically given by the adjoint scalar field generated at the sensor. 
Based on this idea, the cost functional~(\ref{cost_fun_or}) is formulated as 
linear superposition of three different components, 
i.e., the mean and fluctuating components of the adjoint scalar field, and the penalty of the sensor speed. 
This naturally yields the extra-adjoint equation based on which the sensor trajectory is iteratively optimized. 
It is confirmed that the cost functional is monotonically decreased in all cases 
with increasing number of forward-adjoint iterations, and the sensor trajectories 
eventually converges to the optimal ones.

The resultant optimal trajectories were implemented to a moving sensor and 
their performances of scalar source estimation were quantitively evaluated.
It was found that the estimation performance of the single sensor moving 
along the optimal trajectory is drastically improved from that of a stationary sensor,
and almost similar to that of 17 stationary sensors. Systematic optimizations with
the different weighting coefficients in the cost functional imply that 
the ratio $\epsilon$ between the fluctuation and the mean of the sensitivity, i.e., the adjoint
scalar field at the source location, is a primary factor in deciding the estimation performance.
Therefore, we also introduced the new cost functional~(\ref{alter_costfunc}) 
which includes $\epsilon$ only.
It was found that the optimal sensor trajectory under the new cost functional yields
essentially the same performance as that under the original cost functional 
with the best combination of the weighting coefficients for the three components. 
The advantage of the second cost functional is that there is no adjusting parameter.
It was also shown that the optimal sensor trajectory is effective in a wide range
of the pulsating frequencies.

The present results support that maximizing the sensor sensitivity with less fluctuation 
is a promising strategy for optimizing a sensor trajectory. It could be easily extended to
find the optimal arrangement of stationary sensors and also optimization of the trajectories of 
multiple moving sensors. In the present study, it is assumed that the location of the scalar source
is known. The localization of a steady scalar source with stationary sensors based on 
the adjoint-based approach is discussed in a recent study \citep{Wang2019}, and it is interesting to 
consider how these techniques can be further extended for localizing a scalar source with a moving sensor. 

Finally, the present study assumes that the complete information of the spatio-temporal evolution of 
the velocity field is available. Since this scenario is unrealistic, it is important to take into account the uncertainty of 
the velocity field and evaluate its impact on the performance of scalar source estimation. 
Another issue is that the present approach requires iterations of adjoint and extra-adjoint computations, 
and therefore it is still difficult to apply to on-line optimization of a sensor trajectory in real experiments. 
Therefore, it is desirable to extract a simple rule of sensor movement based on the local measurements
from the obtained optimal sensor trajectory. These issues remain to be addressed in future work.


\begin{acknowledgments}
This study is supported by the Japan Science and Technology Council (JST), Strategic
International CollaborativeResearch Program (SICORP), and by the National Science Foundation 
(grant CNS1461870)
\end{acknowledgments}

\newpage

\appendix

\section{ Derivation of the optimization algorithm }

We consider the following cost-functional for optimizing the sensor trajectory:
\begin{equation}\label{j_extented}
J_{1} = - \alpha_{1}\int_{0}^T c^*(\bold{x}^s, t) dt + \alpha_{2}\int_{0}^T 
\bigg( c^{*}(\bold{x}^s, t)-\overline{c^{*}}(\bold{x}^s)\bigg)^{2}dt 
+\alpha_{3}\int_{0}^T \bold{u}^{m}\cdot\bold{u}^{m}dt\,,  
\end{equation}
where $\alpha_{i}$ represent relative importance of each term.
The perturbation of the Hamiltonian function $H$ due to the small change of the sensor trajectory
$\bold{x}^{m}$ is written as 
\begin{equation}\label{Hamiltonian_pert_app}
 H' = J_{1}' + \bigg \langle \theta \bigg \{ \frac{\partial c^{*'} }{\partial t^{*}}
-u_{j}\frac{\partial c^{*'} }{ \partial x_{j} }
-\frac{1}{Pe}\frac{\partial^{2}c^{*'}}{\partial x_{j}^{2} } 
-\frac{ \partial \delta(x_{j}-x^{m}_{j}) }{\partial x^{m}_{j} } x^{m'}_{j} \bigg \} \rangle\,.  
\end{equation}
The first term of the above equation arises from the perturbation of the 
cost-functional, denoted as $H'_{a}$, can be written as
\begin{equation}\label{perturbed_hamiltonianA_or}
H'_{a}=J_{1}' = \langle c^{*'}\bigg \{ -\alpha_{1} + 2 \alpha_{2}\bigg[ c^{*}-\overline{c^{*}}\bigg]\bigg \}\delta(\bold{x}
     -\bold{x}_{s})\rangle + \int_0^T H'_{\mathscr{B},2}dt + \int_0^T H'_{\mathscr{R},2} x^{m'}_{j}dt\,,
\end{equation}
where we introduce the following identities : 
\begin{subequations}\label{HB_HR_B}
\begin{align}
H'_{\mathscr{B},2} & = \frac{d}{d t} \bigg[ 2 \alpha_{3} x^{m'}_{j}\frac{d x^{m}_{j}}{d t}\bigg]\,,\\
H'_{\mathscr{R},2} & = -2\alpha_{3}\frac{d^{2} x^{m}_{j}}{d t^{2}}\,.
\end{align}
\end{subequations} 
The remaining part of eq.~\eqref{Hamiltonian_pert_app}, denoted as $H'_{b}$, can be written as
\begin{eqnarray}\label{perturbed_hamiltonianB_or}
H_{b}' = H'-H'_{a} =\left< {c^*}' \left[ -\frac{\partial \theta }{\partial t^*} + \frac{\partial \left(\theta u_j \right) }{\partial x_j}  
- \left\{ \frac{\partial}{\partial x_j}  \left( \frac{1}{Pe} \frac{\partial \theta}{\partial x_j} \right) \right\}
\right] \right. \nonumber\\
\left. - \frac{\partial \theta}{\partial x_j} \delta\left({x}_j - x^{m}_j\right){ x^{m'}_j }
\right> + \mathscr{B},
\end{eqnarray}
where
\begin{equation}\label{boundary_term}
\mathscr{B} = \left< 
\frac{\partial \theta {c^*}' }{\partial t^*}
+ \frac{\partial}{\partial x_j}  \left\{ - \theta {c^*}' u_j + \frac{1}{Pe} {c^*}' 
\frac{\partial \theta}{\partial x_j} - \frac{1}{Pe} \theta \frac{\partial {c^*}'}{\partial x_j} 
+ \theta  \delta\left({x}_j - x^{m}_j \right) \right\}
\right>.
\end{equation}

Substituting eqs.~\eqref{perturbed_hamiltonianA_or}, \eqref{perturbed_hamiltonianB_or} 
into \eqref{Hamiltonian_pert_app} and introducing a new time coordinate $t^{*}=T-t$ 
yield the following expression:
\begin{equation}\label{perturbed_hamiltonian_aux_app}
\begin{split}
H' & = \langle c^{*'}\bigg[ \frac{\partial \theta}{\partial t} + \frac{\partial (\theta u_{j})}{\partial x_{j}}
-\bigg\{ \frac{\partial }{\partial x_{j} }(\frac{1}{Pe}\frac{\partial \theta }{\partial x_{j} })\bigg\}
+ \bigg\{-\alpha_{1} + 2 \alpha_{2}\bigg[c^{*}-\overline{c^{*}}\bigg]\bigg\} \delta(\bold{x}-\bold{x}^{s})\bigg]\rangle  \\
& -\langle \frac{\partial \theta }{\partial x_{j}}\delta(x_{j}-x^{m}_{j})x^{m'}_{j}\rangle - 
 2 \alpha_{3} \int \frac{ d^{2}x^{m}_{j} }{ dt^{2} }x^{m'}_{j}dt  + \mathscr{B^{*}}\,,
\end{split}
\end{equation}
where the boundary term $\mathscr{B^{*}}$ is given by
\begin{equation}\label{boundary_term}
\mathscr{B^{*}} = \mathscr{B} + 2\alpha_{3} \int_{0}^{T} \frac{d}{dt}\bigg[ x^{m'}_{j}\frac{d x^{m}_{j} }{ dt} \bigg]  \,. 
\end{equation}

In order to remove the first term on the right-hand-side of Eq.~\eqref{perturbed_hamiltonian_aux_app}, 
we impose the following equation for $\theta$:
\begin{equation}\label{theta_evol_eq_app}
\frac{\partial \theta}{\partial t} + \frac{\partial (\theta u_{j})}{\partial x_{j}}
=\bigg\{ \frac{\partial }{\partial x_{j} }(\frac{1}{Pe}\frac{\partial \theta }{\partial x_{j} })\bigg\}
+ \bigg\{\alpha_{1} - 2 \alpha_{2}\bigg[c^{*}-\overline{c^{*}}\bigg]\bigg\} \delta(\bold{x}-\bold{x}^{s}),
\end{equation}
which describes the spatial and temporal evolution of $\theta$. We further assume that the
above equation is  accompanied with the following set of boundary conditions:
\begin{equation}\label{theta_bc_app}
\theta(\bold{x},t=0)=0,\quad  \frac{\partial \theta }{\partial x_{j}}n_{j}=0,\quad \text{at}\,
\partial \Omega.
\end{equation}
so that the boundary term $B^*$ becomes zero. 
Substituting Eqs.~(\ref{theta_evol_eq_app}),~(\ref{theta_bc_app}) into Eq.~(\ref{perturbed_hamiltonian_aux_app})
results in
\begin{eqnarray}\label{simplified_Hvar_app}
   H'=& \left< -\frac{\partial \theta}{\partial x_j} \delta\left({x}_j - {x}^{m}_j\right) x^{m'}_j \right>- 
 2 \alpha_{3} \int \frac{ d^{2} x^{m}_{j} }{ dt^{2} } x^{m'}_{j}dt \nonumber\\
 =& \int_{0}^{T}  -\bigg[ \left( \frac{\partial \theta}{\partial x_j} \right)_{ \bold{x}^{m} }
+2 \alpha_{3}\frac{ d^{2}x^{m}_{j} }{ dt^{2} }\bigg] x^{m'}_j dt\,.
\end{eqnarray}
Equation~(\ref{simplified_Hvar_app}) indicates that $H'$ is always negative 
by updating the sensor trajectory based on the following formula:
\begin{equation}\label{update_exp_app}
x^{m'}_{j}(t)\bigg( \equiv x^{m,n+1}_{j}(t)-x^{m,n}_{j}(t) \bigg)=\alpha^{n}
\bigg[ \left( \frac{\partial \theta}{\partial x_j} \right)_{ \bold{x}^{m} }
+2 \alpha_{3}\frac{ d^{2} x^{m}_{j} }{ dt^{2} }\bigg]^{n},
\end{equation}
where the second superscript of $n$ indicates a iteration step and 
$\alpha^n$ is a positive coefficient determining the amount of the update
in the $n$-th iteration step.

\section{ Derivation of optimization algorithm for the cost-functional $J_2$ }

We seek to find the optimal sensor trajectory to minimize the following cost functional:
\begin{equation}\label{costfunc_app}
J_{2} = \frac{ c^{*}_{rms}(\bold{x}^{s}) }{ \overline{c^{*}(\bold{x}^{s})} }=
\frac{ \bigg\{  \frac{1}{T}\int_{0}^{T} [c^{*}-\overline{c^{*}}]^{2}
\delta(\bold{x}-\bold{x}^{s})dt \bigg \}^{1/2} }{ \frac{1}{T}\int_{0}^{T}
c^{*}\delta(\bold{x}-\bold{x}^{s})dt }\,.
\end{equation}

For convenience, we introduce the following identities:
\begin{equation}
\begin{split}
w & = c^{*}_{rms}(\bold{x}^{s}) = \bigg \{  \frac{1}{T}\int_{0}^{T} [c^{*}-\overline{c^{*}}]^{2}
\delta(\bold{x}-\bold{x}^{s})dt \bigg \}^{1/2} \,, \\
g & = \overline{c^{*}(\bold{x}^{s})} = \frac{1}{T}\int_{0}^{T}
c^{*}\delta(\bold{x}-\bold{x}^{s})dt \,.
\end{split}
\end{equation}
The perturbations of $w$ and $g$ with respect to the sensor location 
$\bold{x}^{m}$ are respectively given by
\begin{equation}
\begin{split}
w' & =  \frac{ \langle  c^{*'}[c^* - \overline{c^*}]\delta(\bold{x}-\bold{x}^{s}) \rangle }{ T\, c^{*}_{rms} }  \,,\\
g' & = \frac{1}{T}\langle c^{*'}\delta(\bold{x}-\bold{x}^{s}) \rangle \,.
\end{split}
\end{equation}
The perturbation of the cost functional~\eqref{costfunc_app} can be obtained 
by applying the quotient rule as follows:
\begin{equation}\label{costfuncper_app}
J_{2}' = \bigg[  \frac{g\,w'-w\,g'}{g^2} \bigg] 
= \frac{1}{T}\langle c^{*'} \bigg \{  
\frac{ [c^* -\overline{c^*}] }{ \overline{c^*}c^{*}_{rms} } - 
\frac{ c^{*}_{rms} }{ \overline{c^{*}}^{2} }\bigg \}\delta(\bold{x}-\bold{x}^s)\rangle \,.
\end{equation}

Applying the same procedure as described in Appendix A to Eq.~(\eqref{costfuncper_app}) 
leads to the following partial differential equation for $\theta$:
\begin{equation}
\frac{ \partial \theta }{ \partial t } + 
\frac{ \partial (\theta u_{j}) }{ \partial x_{j}  } = \bigg \{ \frac{\partial}{\partial x_{j}}\bigg( \frac{1}{Pe}\frac{\partial \theta }{\partial x_{j} }\bigg) \bigg\} 
+ \bigg \{ \frac{c^{*}_{rms} }{ \overline{c^{*}}^{2} }-\frac{ [c^* -\overline{c^*}] }{ \overline{c^*}c^{*}_{rms} } \bigg \}\delta(\bold{x}-\bold{x}^{s})\,.
\end{equation} 
The boundary conditions and the iterative expression for $\bold{x}^{m}$ remain the same as 
those shown in Eqs.~\eqref{theta_bc_app} and \eqref{update_exp_app}, respectively.

\newpage

\bibliography{mylib4}

\end{document}